\documentclass[11pt,a4paper,twoside]{article}

\usepackage{epsfig}
\usepackage{amssymb}
\usepackage[latin1]{inputenc}
\def\be{\begin{equation}}
\def\ee{\end{equation}}
\def\ba{\begin{array}}
\def\bacc{\begin{array} {cc}}
\def\ea{\end{array}}
\def\bea{\begin{eqnarray}}
\def\eea{\end{eqnarray}}
\def\bd{\begin{displaymath}}
\def\ed{\end{displaymath}}

\def\T{{\scriptscriptstyle T}}
\def\V{{\scriptscriptstyle V}}

\begin{document}

\hspace{12.1cm}DESY 09-009

\hspace{12cm} UAB-FT-661


\begin{center}

{\Large\bf General Perturbations for Braneworld Compactifications 
\\ and the Six Dimensional Case}

\vspace{1cm}

{\large S. L. Parameswaran$^a$\footnote{Email:
susha.louise.parameswaran@desy.de}, S. 
Randjbar-Daemi$^b$\footnote{Email: seif@ictp.trieste.it} 
and
A. Salvio$^c$\footnote{Email: salvio@ifae.es}}\\

\vspace{.6cm}

{\it {$^a$ II Institute for Theoretical Physics, \\University of
Hamburg, DESY Theory Group, Notkestrasse 85, Bldg. 2a, D-22603
Hamburg, Germany }} 

\vspace{.4cm}

{\it {$^b$ International Center for Theoretical Physics, \\Strada
Costiera 11, 34014 Trieste, Italy}}

\vspace{.4cm}

{\it {$^c$ Institut de Th\'eorie des Ph\'enom\`enes Physiques,
\\EPFL, CH-1015 Lausanne, Switzerland\\
and \\
IFAE, Universitat Aut\`{o}noma de Barcelona, \\08193 Bellaterra,
Barcelona, Spain}} 

\end{center}

\vspace{0.5cm}

\begin{abstract}

Our main objective is to study how braneworld models of
higher codimension differ from the 5D case and traditional
Kaluza-Klein compactifications.
We first derive the classical dynamics describing the
physical fluctuations in a wide class of models incorporating gravity,
non-Abelian gauge fields, the dilaton and two-form potential, as well
as 3-brane sources.  Next, we use these results to study
braneworld compactifications in 6D supergravity, focusing on the
bosonic fields in the minimal model; composed of the supergravity-tensor
multiplet and the $U(1)$ gauge multiplet whose flux supports the
compactification.  For unwarped models sourced by positive tension
branes, a harmonic analysis allows us to solve the large, coupled,
differential system completely and obtain the full 4D spin-2,1 and 0
particle spectra, establishing (marginal) stability and a qualitative
behaviour similar to the smooth sphere compactification.  We also find
interesting results for models with negative tension branes; extra
massless Kaluza-Klein vector fields can appear in the spectra, beyond
those expected from the isometries in the internal space.  These
fields imply an enhanced gauge symmetry in the low energy 4D effective
theory obtained by truncating to the massless sector,  which is
explicitly broken as higher modes are excited, until the
full 6D symmetries are restored far above the Kaluza-Klein scale.
Remarkably, the low energy effective theory does not seem to
distinguish between a compactification on a smooth sphere and these
singular, deformed spheres.

\end{abstract}

\newpage

\tableofcontents

\newpage

\section{Introduction}

Almost two decades on, branes are evermore ubiquitous in the models
constructed to understand particle physics and cosmology, with all
their How?'s and Why?'s.  As fundamental objects, they are the
D-branes and NS-branes (or M-branes) of string (or M) theory, but within
a low-energy 
effective field theory description, they are introduced as
braneworlds.  Often these braneworlds are considered as infinitely
thin but finite tension objects, like for their more fundamental cousins,
although sometimes 
it proves necessary to resolve their structure by
adding some thickness.

A codimension one brane necessarily forms a boundary in the bulk
space, since there is no path which can lead from one side to the
other without traversing the brane.  The gravitational backreaction
of these objects is well understood; whilst the metric is continuous
across the brane, its first derivative can have finite discontinuities.
Branes with more than one transverse dimension are qualitatively
different, and much harder, due to their sourcing of singularities in the
transverse space.  Still, codimension two branes can also be handled
with some control; they backreact on the geometry in such a way as to
produce relatively mild conical singularities.

The construction of solutions sourced by branes, with up to two codimensions,
in various field theory models is by now a well-developed art.  In 5D,
the archetype is of course the construction of Randall and Sundrum
\cite{Randall:1999ee, RSb}.
In 6D, we take the general warped braneworld compactifications (``conical-GGP
solutions'') of 6D N=1 gauged 
supergravity \cite{Nishino:1984gk} 
found in \cite{Gibbons:2003di, 8author} as
representative.  These solutions additionally invoke fluxes, which are
also playing a dominant role in string compactifications today, and
indeed models with two extra dimensions are the simplest in which flux
compactifications can be studied.
Having established the solutions, we can begin to ask 
about their physics: Are they stable to small perturbations?  What are
the symmetries and particle content of the low energy effective field
theory?  Is it chiral?  What are the modifications to 4D Einsteinian
gravity?  What would be the effective vacuum energy measured by a 4D
observer?  What role do the branes play in these  
and other phenomena? And so on.

The first step towards answering these questions is to analyze the
classical spectra of small fluctuations around the solution.  A number of such
studies have been made recently for the conical-GGP solutions.  
In \cite{Parameswaran:2006db} we worked out the spectra for certain
4D gauge fields and fermions present in the model and no tachyons or
ghosts were found amongst them. A similar (marginal) stability was
found in \cite{Burgess:2006ds}, where the axially symmetric modes for
some of the scalar perturbations were calculated. The spectrum for
the gravitino has also been analyzed in \cite{gravitino}. In
\cite{Parameswaran:2007cb}, meanwhile, we studied the tachyonic
instabilities that can arise from the non-axially symmetric, 4D scalar
fluctuations 
descending from 6D gauge fields, and charged under the 
background fluxes\footnote{The end point of this instability
  is studied in \cite{endpoint}.}.  Whether a given model with a given flux
suffers from this instability turns out to depend on the tensions of
the branes present.

We now intend to complete the spectral analysis for the bosonic
fluctuations about the braneworld solutions of 6D supergravity.  Our
particular focus in this paper is on the so-called Salam-Sezgin sector -- that 
arising from the supergravity-tensor multiplet and the $U(1)$ gauge multiplet
in which the background monopole lies -- which was partially treated
in \cite{Lee:2006ge,Burgess:2006ds}.  The remaining sectors have been
completed elsewhere \cite{Parameswaran:2006db,Parameswaran:2007cb}.  We will  
calculate the corresponding spectra for the 4D spin-2 and -- for
unwarped backgrounds -- spin-1 and spin-0 fields.  The model that we
are studying is complicated, and technically difficult.  However, this
goes hand in hand with its advantage of generality, and indeed the
results for several simpler scenarios can be extracted from our work
at its various stages.  

Our approach will be that established in \cite{RandjbarDaemi:2002pq}, where a
formalism was developed to analyze the spectra of small perturbations
about arbitrary solutions of Einstein, Yang-Mills and scalar systems.
The first part of this paper can be considered as a generalization of that
work, where we now include the presence of thin source 3-branes and extra
bulk fields that are generically present in supergravity theories; the dilaton
and anti-symmetric two-form potential.  With little extra 
cost, we actually keep the number of dimensions transverse to the brane
general.

We first derive
the general form of the bilinear action that 
describes the behaviour of small fluctuations.  For codimension-two or
higher, we include fluctuations of the brane positions in the
transverse directions, the so-called ``branons''.  We then apply the
light-cone gauge (for bulk fields) and static gauge (for branons) to
restrict to physical degrees of freedom, and 
decouple the dynamics for the spin-2, -1 and -0 fluctuations. 
The gauge-fixed bilinear action thus obtained provides the starting
point to calculate the Kaluza-Klein (KK) spectra 
for the conical-GGP solutions, as well as, for example, the 5D
Randall-Sundrum models and the 
non-supersymmetric Einstein-Yang Mills(-dilaton) model in any
dimension.  

In the second part of this paper, we use these general equations to
study the behaviour of braneworld models in 6D (along the way also
recover some of well-known aspects of the 5D scenarios).  Here, since
we include the backreaction of the branes, the dynamics of the branons
are not well-defined\footnote{Indeed, the behaviour of the branons is usually
considered under the probe brane approximation, in which the brane tension
is much smaller than the bulk gravitational scale, so that the
backreaction can be neglected \cite{Sundrum:1998ns,branons}.}.
Therefore, to study the spin-0 sector, we choose to 
truncate the branons by {\it e.g.} 
placing the branes at orbifold fixed points, or taking the brane
tensions to be very large making the branes rigid within our
range of validity.  Meanwhile, the 
conical singularities in the curvature that are induced by the
codimension two branes do not
prevent us from understanding the behaviour of the bulk fluctuations. 

We are able
to derive the spectrum for the 4D spin-2 fields in the model's full
warped generality.  The spin-1 and spin-0 sectors present large
coupled differential systems, and by finding a set of harmonics on the
2D internal space ( the ``rugbyball''), we are also able to solve
these systems analytically for the unwarped case.  In this way, we
obtain {\em all} the 4D modes for unwarped compactifications with
positive tension brane sources, and qualitatively, we observe the same
behaviour as in the smooth
sphere compactification without branes -- including marginal
stability.  

In the presence of negative 
tension codimension-two branes, meanwhile, the physics can surprise.
Here, despite the fact that brane sources clearly break the $SU(2)$ isometries
of the sphere to $U(1)$, three massless spin-1 fields\footnote{In addition to any massless
  gauge fields arising from unbroken higher dimensional gauge
  symmetries.} can be found
amongst the KK spectra for special values of the conical deficit
angle.   These special deficit angles, $\delta=-2\pi, -4\pi, \dots$, allow three Killing vectors to be well-defined
everywhere outside the branes, although only one of them can be
globally integrated to an isometry.  

Whether or not the massless 
vectors are gauge bosons of an enhanced gauge symmetry in the 4D
theory can be understood by going beyond bilinear order and
considering the interaction terms.  We find the presence of
KK modes that are not in well-defined representations of the
$SU(2)$ generated by the Killing vectors, and therefore the full 4D
theory does not enjoy an $SU(2)$ gauge symmetry.  For this
reason, we do not expect the classical masslessness of the vector fields to
survive quantum corrections.   Meanwhile, all our bosonic massless modes
do fall into 
well-defined $SU(2)$ representations, and therefore we argue that the
classical low energy 4D effective field theory -- obtained by truncating to the
massless sector -- does enjoy an enhanced KK gauge symmetry
beyond the isometries!  Moreover,
it appears that the low energy theory does not distinguish between
compactifications on the smooth sphere and these singular, deformed spheres.

Let us now give an outline for the remainder of the paper.  The first
part presents a rather general analysis that determines the dynamics
of perturbations in braneworld compactifications.  In the 
next section, we introduce the model (both theories and background
solutions) and discuss the scenarios to which our analysis can be
applied.  In Section 3, we introduce the 
perturbations about the background, obtain the bilinear action that
describes their dynamics, and discuss the local symmetries of this
action.  In Section 4, we use these symmetries to fix to the ``light
cone static gauge'', and give the bilinear action in this gauge, in
which the different spin sectors decouple.  

Then begins the second part, which uses the previous results to study
the 4D fields that emerge in various scenarios.  In Section 5, our
main interest is in the braneworld solutions of 6D supergravity, but
we also discuss a non-supersymmetric 6D model and the 5D
Randall-Sundrum models.  In the main text we present the KK spectra
for spin-2 and spin-1 fields and identify the massless spin-0 
fields; the complete spin-0 sector can be found 
in the appendices.  Finally, we understand in detail the physical
significance of the extra massless 4D vector modes that can appear in the
spectra, and the gauge invariance
that emerges in the 4D theory. 

We summarise our results in Section 6, before concluding in Section 7.  

\section{The Model}

We begin with the definition of our model.  The main focus of the
present paper will be a class of bosonic 6D field theories with thin
codimension-two branes. In particular we are interested in the bosonic
part of 6D N=1 gauged supergravity \cite{Nishino:1984gk}. However,
throughout the article we shall keep a general space-time dimension
$D$ as far as possible, and certain truncations of the field content
allow our analysis to be applied to several different scenarios,
including the non-supersymmetric
Einstein-Yang-Mills theory or the Randall-Sundrum Model.

\subsection{Field content}

The basic ingredients of our model are the higher dimensional metric
$G_{MN}$, where the space-time indices run over $M, N,...=0, ...,
D-1$, and the gauge field 
$\mathcal{A}_M$ of a compact Lie group $\mathcal{G}$. These are bulk
fields in the sense that they depend on all the space-time coordinates
$X^M$. 

We also want to consider a certain number $\mathcal{N}$ of 3-branes
embedded in the $D$-dimensional space time. To do so we introduce,
following Ref. \cite{Sundrum:1998sj}, $\mathcal{N}$ functions
$Y^M_k(x_k), \, k=1,...,\mathcal{N}$, which represent the positions of
the branes in the $D$-dimensional space time. The $x_k$ represent
the 4D coordinates on the brane, $x_k= \left\{x_k^{\alpha}\right\}$,
where $\alpha, \beta,...$ are the 4D indices. Not all the space-time
components of $Y^M_k(x_k)$ are physical degrees of freedom: 4
space-time components for each $k$ can be gauged away by using the 4D
(general) coordinate transformation invariance acting on $x_k$
\cite{Sundrum:1998sj}, as we will explicitly do in Subsection
\ref{gauge-fixing}. We consider $Y^M_k(x_k)$ to be a brane field because
it depends only on a 4D world-volume coordinate. These fields are
important to introduce the branes in a covariant way, and indeed we
can construct the induced metrics on the branes by means of 
\be g_{\alpha \beta}^k=
G_{MN}(Y_k(x_k))\partial_{\alpha}Y^M_k(x_k)\partial_{\beta}Y^M(x_k) \,
.
\label{brane-metric}\ee 

In order to complete the bosonic part of the 6D
supergravity, one should add other
bulk fields in addition to $G_{MN}$ and $\mathcal{A}_M$, that is a
dilaton $\phi$ and a 2-form field $B_{MN}$, which emerge from the
graviton multiplet and an antisymmetric tensor multiplet
\cite{Nishino:1984gk}.  We will refer to $B_{MN}$ as the Kalb-Ramond
field. Moreover, 
concerning the 6D supergravity, we shall assume that $\mathcal{G}$ is
a product of simple groups that include a $U(1)_R$ gauged R-symmetry. 
In general one can also add some hypermultiplets
\cite{Nishino:1984gk}, which turn out to be important to cancel gauge
and gravitational anomalies \cite{seifanomaly,al-anomaly}. In the
bosonic sector this leads to additional scalar fields $\Phi^{\alpha}$
(hyperscalars) in some representation of $\mathcal{G}$; however, from
now on we set $\Phi^{\alpha}=0$. We do so because we are interested in
the linear perturbations which mix with the 
$D$-dimensional gravitational fluctuations
$h_{MN}$: indeed, for the class of backgrounds we are interested in
(see Subsection \ref{EOM}), the $\Phi^{\alpha}$ decouple from
$h_{MN}$.  Their inclusion should be straightforward. 

Therefore the bulk and the brane field contents that we consider are
respectively:

\be \left\{G_{MN}, \mathcal{A}_M, \phi, B_{MN} \right\} \quad
\mbox{and} \quad \left\{Y^M_k(x_k), ...\right\}.\label{fields}\ee 
The dots in the second set of (\ref{fields}) represent additional
brane fields that we can always introduce, but which are not required by
general covariance; for example they can be the fields of the Standard
Model (SM).  

\subsection{The action}
We split the action functional $S$ into the bulk action $S_B$, which
depends only on the bulk fields, and the brane action $S_b$ that is a
functional of the brane fields as well. 

The bulk action is\footnote{We choose signature $(-,+,..,+)$, and
  define  $R_{MN\phantom{R}S}^{\phantom{MN}R}=\partial_M \Gamma_{NS}^R
  -\partial_N \Gamma_{MS}^R 
+ \Gamma_{MP}^R \Gamma_{NS}^P -\Gamma_{NP}^R \Gamma_{MS}^P$ and
$R_{MN}=R_{PM\phantom{P}N}^{\phantom{PM}P}$.}\cite{Nishino:1984gk} 
\be S_B=\int d^DX
\sqrt{-G}\left\{\frac{1}{\kappa^2}\left[R-\frac{1}{4}\left(\partial
      \phi\right)^2\right] 
-\frac{1}{4}e^{\phi/2}F^2
-\frac{\kappa^2}{48}e^{\phi}H_{MNP} H^{MNP} -\mathcal{V}(\phi)\right\},
\label{SB} \ee
where $G$ is the determinant of $G_{MN}$ and $\kappa$ is the
$D$-dimensional Planck  
scale; also\footnote{A trace overall is understood when we write a
  product of Lie algebra valued objects: e.g. in Eq. (\ref{SB}) 
$F^2\equiv \mbox{Tr}\left(F^2\right)$.} $F^2\equiv F_{MN} F^{MN}$ and
$\left(\partial \phi\right)^2\equiv
\partial_M\phi\,\partial^M\phi$. The explicit expression for the gauge
field strength $F_{MN}$ is\footnote{We define the cross-product as 
$({\cal A}_M \times {\cal A}_N)^{ I} = f^{ I J K} 
{\cal A}_M^{ J}{\cal A}_N^{ K}$, with $f^{ I J
K}$ the structure constants of ${\mathcal G}$:
$\left[T^I,T^J\right]=if^{IJK}T^K$, where $T^I$ are the generators of
$\mathcal{G}$.} 
\be F_{MN}=\partial_M\mathcal{A}_N
-\partial_N\mathcal{A}_M+g\mathcal{A}_M\times \mathcal{A}_N,\ee 
where $g$ is the gauge
coupling, which in fact represents a collection of
independent gauge couplings including that of the $U(1)_R$
subgroup, $g_1$.
$H_{MNP}$ is the Kalb-Ramond field strength, which contains a Chern-Simons
coupling as follows \cite{Nishino:1986dc}:
\be H_{MNP} = \partial_M B_{NP} +
F_{MN}{\cal A}_P - \frac{g}{3}{\cal A}_M\left({\cal A}_N
\times {\cal A}_P \right) + \,\rm{2 \,\, cyclic \,\, perms} \,. \label{HKR}\ee
The function $\mathcal{V}(\phi)$ is the dilaton potential. In the
supersymmetric model this is fixed to be $\mathcal{V}(\phi)=8\,g_1^2
\,e^{-\phi/2}/\kappa^4$. 

Meanwhile, we consider the following 3-brane action 
\be 
S_b= \sum_k\left(-T_k\int d^4x_k\sqrt{-g^k}\right)\equiv -T\int
d^4x\sqrt{-g}, \label{Sb} 
\ee 
where $g^k$ is the determinant of (\ref{brane-metric}) and  $T_k$ are
the tensions of the branes. From now on (unless otherwise
stated) we suppress the index $k$, as we have done on the right hand side of
(\ref{Sb}).  
The reader may have noticed that we have not introduced the
Gibbons-Hawking boundary term, which is generically necessary to treat
codimension one branes \cite{GibbonsHawking}.  Indeed, we shall apply
our analysis only to those codimension one models whose branes are placed on
orbifold fixed points, in which case the Gibbons-Hawking boundary term
is not present \cite{intervalvsorb}.

We can summarise by saying
that our analysis will apply to the following two types of models:
\begin{enumerate}
 \item 6D N=1 gauged supergravity.
\item Einstein-Yang-Mills theories, with a dilaton or
  cosmological constant  
$\Lambda$, for a general
  space-time dimension.  
\end{enumerate}
 The second case includes, for example, the RS models
 \cite{Randall:1999ee,RSb} or the non-supersymmetric 6D
Einstein-Yang-Mills-$\Lambda$ (EYM$\Lambda$) model \cite{Sundrum:1998ns,
  Carroll:2003db}.  They can be obtained by simply 
fixing the
appropriate dimension and setting $H_{MNP}=0$,
  $\phi=0$ and 
$\mathcal{V}(0)=\Lambda$. Even if our main interest is in models of
Type 1 we will also 
consider the second class for several reasons.  In this way, we will
see that our results can be applied in quite general contexts, and it
will also provide interesting additional ways to check our formulae.
Moreover, in the future it should help us to figure out the role of
supersymmetry in 
the linear perturbations.

Finally, it is important to note that the actions $S_B$ and $S_b$ are
invariant with respect to both the $D$-dimensional and the 4D coordinate
transformations (acting respectively on $X^M$ and $x^{\alpha}$). We
will discuss the local symmetries of the present model and an explicit
gauge fixing for the linear perturbations in Subsections
\ref{local-symmetries} and \ref{gauge-fixing}.

\subsection{The equations of motion (EOMs) and solutions} \label{EOM}

The EOMs that follow from the variation of the action $S_B+S_b$ are:
\bea R^{MN}-\frac{1}{2}G^{MN}R&=&
\frac{\kappa^2}{2}\left\{e^{\phi/2}\left(F^M_{\,\,\,\,\,P}F^{NP}-\frac{1}{4}G^{MN}F^2\right)+   
\frac{1}{2\kappa^2}\partial^M\phi \,\partial^N\phi \right. \nonumber
\\&&\left.-G^{MN}\left[\frac{1}{4\kappa^2} \left(\partial\phi\right)^2
    +\mathcal{V}(\phi)\right] 
\right\} -T\kappa^2 \, \mathcal{B}^{MN}, \label{GMN-EOM} \\
D_N\left( e^{\phi/2} F^{NM}\right)&=&0, \label{A-EOM}\\
\frac{1}{2\kappa^2}D^2\phi &=& \frac{\partial \mathcal{V}}{\partial
  \phi}(\phi)+\frac{1}{8}e^{\phi/2} F^2,\label{phi-EOM}\\ 
\frac{1}{\sqrt{-g}} \partial_{\alpha}\left( \sqrt{-g}
  \,G_{MN}\partial^{\alpha}Y^N\right) &=& \frac{1}{2}\,\partial_M
G_{NP} \, \,\partial Y^N \cdot \partial Y^P,\label{Y-EOM} 
\eea
where we have fixed $H_{MNP}=0$, since our interest shall be in
backgrounds that enjoy 4D Poincar\'e invariance.  Moreover, in
Eq. (\ref{phi-EOM}) and (\ref{Y-EOM}) we have introduced the 
notation $D^2\phi \equiv D_M D^M \phi$, where $D_M$ is the covariant
derivative, and $\partial Y^{M}\cdot \partial Y^N \equiv
\partial_{\alpha} Y^{M} \partial^{\alpha} Y^N$.  Recall also that we
have suppressed the index $k$ on $Y^M_k$, which labels each of the branes. 
The last term in (\ref{GMN-EOM}) represents the brane contribution to
the Einstein equations, where $\mathcal{B}^{MN}$ is defined by 
\be \mathcal{B}^{MN}(X)\equiv \frac{1}{2}\int d^4 x \sqrt{g/G} \,\,
\delta (X-Y(x))\,\,\partial Y^{M}\cdot \partial Y^N; \label{B-def}\ee  
we note that the bulk quantity $G$ in (\ref{B-def}) is computed at the
position of the brane ($G=G(Y)$) because of the presence of the
$D$-dimensional delta function $\delta (X-Y(x))$. 
Furthermore, since Eqs. (\ref{Y-EOM}) come from the variation of the brane
action with respect to $Y^M$, there the bulk fields $G_{MN}$ and
$\partial_M G_{NP}$ are computed at the brane position ($G_{MN}=G_{MN}
(Y)$ and $\partial_M G_{NP}=\partial_M G_{NP} (Y)$). 

In the present paper we will focus mainly on the following ansatz
solution to (\ref{GMN-EOM})-(\ref{Y-EOM}): 
\bea Y^{\mu}&=& x^{\mu},\label{static-background}\\
Y^{\underline{m}} &=& \mbox{constant}\,, \label{constant-Ym} \\ 
ds^2&=& e^{A(\rho)} \eta_{\mu \nu}dx^{\mu} dx^{\nu} + d\rho^2 +
e^{B(\rho)} K_{mn}(y) dy^{m}dy^{n}\,,  \label{background-metric}\\ 
\mathcal{A}&=& \mathcal{A}_m(\rho, y) dy^m\,, \label{background-gauge}\\ 
\phi &=& \phi (\rho), \label{background-dilaton}\\
H_{MNP}&=&0 \,, \label{background-H}
\eea
where $\mu = 0,1,2,3$, $m=5,...,4+D_2$, $\underline{m}= (\rho,m)$
(we have $D=5+D_2$) and $y^m$ and  $K_{mn}$ are respectively the
coordinate and  the metric on the $D_2$-dimensional space. 
Eq. (\ref{static-background}) is not really an assumption because we can
always use the 4D general coordinate invariance on the branes to set
(\ref{static-background}). Eq. (\ref{constant-Ym}) is instead a non
trivial assumption. Moreover, in Eqs
(\ref{background-metric})-(\ref{background-H}) we are assuming that
the bulk field background has a 4D Poincar\'e invariance and that the
functions $A$, 
$B$ and $\phi$ depend only on the coordinate $\rho$. We will also
assume $\mathcal{A}$ to lie in the Cartan subalgebra of
Lie$(\mathcal{G})$. 

One of the simplest models that can be described by this set up is the
Randall-Sundrum (RS) model \cite{Randall:1999ee}, where we have $D=5$,
$\phi=0$ and $\mathcal{A}=0$ and the internal space is $S^1/Z_2$ with
two branes 
on the fixed points of $Z_2$, say at $\rho=0$ and $\rho=\pi r_c$.
The explicit form of the solution is given by
\be  A= -2k|\rho|, \quad Y^{\rho}_1=0, \quad Y^{\rho}_2=\pi
r_c\,, \label{RS-solution}\ee 
where $k$ is a positive constant. The object $|\rho|$ in
(\ref{RS-solution}) is equal to the absolute value of $\rho$ in the
region $-\pi r_c< \rho < \pi r_c$ and its value anywhere else is
obtained by periodicity. In order for (\ref{RS-solution}) to 
be a solution one needs  $T_1=-T_2=12k/\kappa^2$ and $\Lambda= -12
k^2/\kappa^2$.  In Section \ref{Applications}, we shall use this very
well-known solution to check the result given in Section \ref{LCG}.

However, in this paper our main interest lies in the analysis of a
class of solutions found by Gibbons, G\"uven and Pope (GGP)
\cite{Gibbons:2003di} to the 6D supergravity: the general set of
warped solutions with 4D Poincar\'e symmetry, and axial symmetry in
the transverse 
dimensions. Here we give only a subset of this general class, namely
that which contains singularities no worse than conical and therefore
can be sourced by brane terms of the form (\ref{Sb}).

To give the explicit expression of the conical-GGP solutions, it turns
out to be useful to introduce the following radial coordinate
\cite{Parameswaran:2006db} 
\be u(\rho)\equiv \int_0^{\rho}d\rho' e^{-A(\rho')/2},
\label{theta}\ee
whose range is $0\leq u \leq \overline{u}\equiv \pi r_0/2$. In this
frame the metric reads 
\be ds^2=e^{A(u)}\left(\eta_{\mu \nu}dx^{\mu}dx^{\nu}+du^2\right)
+ e^{B(u)} \, \frac{r_0^2}{4} \, d\varphi^2 \, .\ee
The explicit conical-GGP solutions\footnote{The coordinate
$u$ is related to the coordinate $r$ in \cite{Gibbons:2003di} by
$r=r_0\cot(u/r_0)$.} are then the following particular case of the ansatz
(\ref{static-background})-(\ref{background-H}) \cite{Gibbons:2003di}:
\bea e^A&=&e^{\phi/2}=\sqrt{\frac{f_1}{f_0}}, \quad e^B=4\,\alpha^2
e^A\frac{\cot^2(u/r_0)}{f_1^2},\nonumber\\
\mathcal{A}&=&-\frac{4\alpha}{q\kappa f_1}\, Q\,
d\varphi,\label{GGPsolution}\eea
where $q$ and $\alpha$ are generic real numbers and $Q$ is a generator
of a $U(1)$ subgroup of a simple factor of $\mathcal{G}$, satisfying
Tr$\left(Q^2\right)=1$.  Also,
\be f_0\equiv 1+\cot^2\left(\frac{u}{r_0}\right), \quad f_1 \equiv
1+\frac{r_0^2}{r_1^2}\cot^2\left(\frac{u}{r_0}\right), \label{GGPsolution2}\ee
with $r_0^2\equiv \kappa ^2/(2g_1^2)$ and  $r_1^2\equiv 8/q^2$.

This solution is supported by two branes located at $u = 0$ and  $u =
\overline{u}$.  Indeed, as $u \rightarrow 0$ or $u \rightarrow
\overline{u}$, the metric tends to that of a cone, with respective
deficit angles   
\be \delta = 2\pi \left(1-|\alpha| \, \frac{r_1^2}{r_0^2}\right)
\quad \mbox{and}\quad
\overline{\delta}=2\pi\left(1-|\alpha| \right) \, , \label{deltadeltabar}\ee
and corresponding delta-function behaviours in the Ricci scalar.  We
will take $\alpha \geq 0$ without loss of generality. The tensions of the
two branes $T$ and $\overline{T}$ are related to the deficit angle as 
follows \cite{chenlutyponton}: 
\be T=2\delta/\kappa^2
\quad \mbox{and} \quad
\overline{T}=2\overline{\delta}/\kappa^2.\label{Tdelta}\ee 
Unlike the RS solution, here the warp factor $e^{A}$ is smooth on the brane
positions $u=0$ and $u=\overline{u}$. In particular we have 
\be e^A \,\stackrel{u\rightarrow 0, \overline{u}}{\rightarrow}\,
\mbox{constant} \neq 0,\quad 
\partial_u e^A\,\stackrel{u\rightarrow 0,\overline{u}}{\rightarrow}\,
0. \label{asymptotics-A}\ee 

By using (\ref{asymptotics-A}), (\ref{static-background}) and
(\ref{constant-Ym}), it is also easy to check that the conical-GGP
configuration satisfies the $Y$-equations (\ref{Y-EOM}) in addition to
the bulk EOMs (\ref{GMN-EOM})-(\ref{phi-EOM}). 

The expression for the gauge field background in Eq. (\ref{GGPsolution}) is
well-defined in the limit $u\rightarrow 0$, but not as $u \rightarrow
\overline{u}$.  We should therefore use a different patch to describe the
$u={\overline{u}}$ brane, and this must be related to the patch including the
$u=0$ brane by a single-valued gauge transformation.  This leads to a Dirac
quantization condition, which for a field interacting with
$\mathcal{A}$ through a charge $e$ gives
\be -e \, \frac{4\alpha \overline{g}}{\kappa q}=- e \,\alpha
\frac{r_1}{r_0} \frac{\overline{g}}{g_1} = 
N \, , \label{DiracQ}\ee
where $N$ is an integer that is called monopole number and
${\overline g}$ is the gauge coupling constant corresponding to
the background gauge field. For example, if ${\mathcal A}$ lies in
$U(1)_R$, then ${\overline g}=g_1$. The charge $e$ can be computed
once we have selected 
the background gauge group, since it is an eigenvalue of the generator $Q$.
Also, note that the internal space corresponding to Solutions
(\ref{GGPsolution}) has an $S^2$ topology (its Euler number
equals 2).

Finally, we observe that one can obtain the unwarped ``rugbyball''
compactification 
\cite{Carroll:2003db} simply by setting $r_0=r_1$. 
In this case the metric is
\be ds^2 =\eta_{\mu \nu}dx^{\mu} dx^{\nu}
+\frac{r_0^2}{4}\left(d\theta^2 + \alpha^2 \sin^2\theta \,\,d\varphi^2 \right),
\label{rugbyball-metric}\ee 
where $\theta \equiv 2 u /r_0$, and the
background value of the dilaton is zero;
therefore this is a solution also to the non-supersymmetric 6D
EYM$\Lambda$ model.  For $\alpha <1$ the deficit angle is positive.  The geometry is also well-defined when $\alpha>1$ and the deficit angle is negative; we name these spaces ``saddle-spheres'' (see \cite{Parameswaran:2007cb} for a detailed discussion on their properties). Moreover, we can smoothly retrieve the
sphere compactification (with radius $r_0/2$) by taking $\alpha
= 1$ in addition to $r_0 =r_1$. 

\section{General Perturbations} \label{Perturbations}

The main purpose of this paper is to study the linear perturbations in
the above models.  We therefore perturb the fields in (\ref{fields})
as follows: 
\bea && G_{MN} \rightarrow G_{MN} + h_{MN}, \quad
\mathcal{A}_M\rightarrow \mathcal{A}_M +V_{M}, \quad  
\phi \rightarrow \phi +\tau, \nonumber \\ && B_{MN}  \rightarrow
B_{MN} + b_{MN}, \quad 
 Y^M \rightarrow Y^M +\xi^M. \label{perturbations}\eea
The first terms in the right hand sides of (\ref{perturbations})
represent the background quantities of the corresponding fields. 
In fact, it is useful to introduce another 2-form field $V_{MN}$ in
order to describe the fluctuations of the Kalb-Ramond field. This can
be done as follows. Since $H_{MNP}$ appears only quadratically in
(\ref{SB}), and $H_{MNP}=0$ at the background level due to 4D
Poincar\'e invariance, the linear
approximation (which corresponds to the bilinear level in the action)
involves only the linear perturbation of $H_{MNP}$, that we denote
with\footnote{Since the background $H_{MNP}=0$, and
  the background 
  monopole, $\mathcal{A}$, lies in the Cartan subalgebra, we see that
  the exterior derivative acting on the background Kalb-Ramond
  potential $B_2$ must be zero.  Also, $\mathcal{A}\wedge
  \mathcal{A}=0$.} $H^{(1)}_{MNP}$, 
\be H^{(1)}_{MNP} = \left[d \left(b_2 - \mathcal{A} \wedge V \right)
  +2 F\wedge V\right]_{MNP},\label{H-1}  
\ee
where we have used the notation of p-forms and $b_2$ is the
fluctuation in the Kalb-Ramond 2-form, $\mathcal{A}$ and $F$ the
background values of the gauge field and its field strength
respectively and $V$ the perturbation of the gauge field.
We now introduce the 2-form $V_2$ as follows: 
\be V_2 \equiv \kappa \left(b_2 -\mathcal{A} \wedge V\right), \ee
whose components will be denoted by $V_{MN}$. $H^{(1)}_{MNP}$ can now
be expressed in terms of $V_2$ and $V$:  
\be  H^{(1)}_{MNP} = \left( \frac{1}{\kappa}d V_2 +2 \gamma F\wedge
  V\right)_{MNP}, \label{H-linear}\ee 
where we have introduced a new parameter $\gamma$;
for $\gamma=1$ we recover the structure of $H^{(1)}_{MNP}$ required by
the 6D supergravity, whereas for $\gamma=0$ the fluctuations of
$V_{MN}$ are completely decoupled (at the linear level) from the
rest. This will allow us to treat simultaneously the 6D supergravity
and the EYM$\Lambda$ models.

Finally, we note that the fields $\xi^M(x)$ describe the fluctuations
of the brane positions, and as such they
are 4D fields.  

\subsection{Bilinear action}
Here we provide the linearized theory which corresponds to the
bilinear approximation in the action. The bilinear action has been
computed by considering the variation of $S_B+S_b$ under
(\ref{perturbations}) and by keeping only terms up to the quadratic
order\footnote{The EOMs (\ref{GMN-EOM})-(\ref{Y-EOM}) guarantee that
  the linear terms vanish.}. We split it into different contributions as
follows: 
\bea && S(h,h) + S(V,V)+ S(h,V) + S(\tau,\tau) + S(h,\tau) +S(V,\tau)
\nonumber \\&&+ S(V_2,V_2) +S(V, V_2) + S(\xi, \xi ) + S(h,\xi) \,,
\label{split}\eea
where $S(h,h)$ is the bilinear action that depends only on the
fluctuations $h_{MN}$, $S(h,V)$ represents the mixing term between
$h_{MN}$ and $V_M$ and so on. We have $S(h,V_2)=S(\tau, V_2)=0$ as a  
consequence of our background ansatz, for which $H_{MNP}=0$. We give
here the explicit 
expressions for the bilinear action that depend only on the bulk
fields; the dynamics of the $\xi^M$ fields, are explicitly given in
Appendix \ref{general-xi}. 
We find:
\begin{eqnarray}
S(h,h)=\int d^DX \sqrt{-G}\hspace{-0.5cm}&&\left\{\frac{1}{2\kappa^2}
\left[
\left(h^{MN}_{\quad ;M}
-\frac{1}{2}h^{;N}\right)^2
-\frac{1}{2}h^{NP}_{\quad;M}h_{NP}^{\quad ;M} +
\frac{1}{4}h^{;M}h_{;M}- \frac{1}{2}R_1 h^2 
\right]
\right. \nonumber \\&& - \frac{1}{2}h_{PM}h^P _{\,\,\,N}
\left(\frac{1}{2} e^{\phi /2} F^{MR}F^N_{\quad R} +
  \frac{1}{4\kappa^2} 
\partial^M\phi\,\partial^N\phi\right)
\nonumber \\&& -\frac{1}{2}h^{MN}h^{PR}
\left(\frac{1}{\kappa^2}R _{PMNR}-
\frac{1}{2}e^{\phi/2}F_{PM}F_{NR}\right)\nonumber \\&&
\left. -\frac{T}{2}\left[\mathcal{B}^{MN}\left(
      h_{PM}h^{P}_{\,\,\,N}-h \,h_{MN}\right) +\frac{1}{2}
    \mathcal{B}^{MNPR}h_{MN}h_{PR}\right]\right\}~,\label{h-h} 
\end{eqnarray}
where the semicolon denotes the (background) gravitational covariant
derivative, $h\equiv G^{MN}h_{MN}$, $R^{\quad \,\,\, P}_{MN\,\,\,R}$
is the  Riemann tensor for the background metric and we have defined 
\be \frac{2}{\kappa^2} R_1 \equiv \frac{1}{\kappa^2}R-\frac{1}{4}
e^{\phi /2}F^2 -\frac{1}{4 \kappa^2} \left(\partial \phi
\right)^2-\mathcal{V}(\phi)  \ee 
and 
\bea \mathcal{B}^{MNPR}\equiv \int d^4x \,
\sqrt{g/G}\,\,\delta(X-Y(x))\hspace{-0.7cm}&&\left[\frac{1}{2}\left(\partial
    Y^M \cdot 
    \partial Y^N\right) \partial Y^P \cdot \partial Y^R \right.
  \nonumber \\ &&\left.-\left(\partial Y^M \cdot \partial
      Y^P\right)\partial Y^N \cdot 
  \partial Y^R\right]. \eea 
The term proportional to $\mathcal{B}^{MNPR}$ in the last line of
(\ref{h-h}) is the contribution to $S(h,h)$ coming from the  brane
action $S_b$, whereas the term proportional to $\mathcal{B}^{MN}$
comes from the EOMs (\ref{GMN-EOM}), which we have used to write
$S(h,h)$ in the form (\ref{h-h}). Moreover, 
\bea
S(V,V)&=&\int d^DX \sqrt{-G}\left[-\frac{1}{2} e^{\phi/2}\left(D_M V_N
    D^M V^N - D_MV_N D^N V^M\right) \right. \nonumber \\
&&\left. -\frac{\kappa^2}{12} \gamma^2 e^{\phi}
  \left(F_{[MN}V_{P]}\right)\left( 
  F^{[MN}V^{P]}\right)-\frac{1}{2}\overline{g} e^{\phi /2} F^{MN} V_M\times
  V_N\right], \label{V-V}  
\\
 S(h,V)&=& - \int d^DX \sqrt{-G}\, e^{\phi/2}\left(D^M V^N-D^N V^M \right)
 \left(\frac{1}{4}h \, F_{MN} + h_{P N} F^P_{\,\,\,M} \right), \label{h-V}\\
 S(\tau,\tau) &=& -\int d^DX \sqrt{-G}
 \left[\frac{1}{4\kappa^2}\left(\partial \tau\right)^2 
+\frac{1}{2} \frac{\partial^2\mathcal{V}}{\partial \phi^2} \, \tau^2
+\frac{1}{32}e^{\phi /2} F^2 \, \tau^2\right], \label{d-d}\\ 
S(h,\tau)&=& \int d^DX \sqrt{-G} \left[\frac{1}{2\kappa^2} \partial_M
  \tau\, \partial_N \phi \,\left(h^{MN} -\frac{1}{2} G^{MN} \,h
  \right) -\frac{1}{2} \frac{\partial \mathcal{V}}{\partial \phi}\, h
  \, \tau \right. \nonumber \\ &&\left. +\frac{1}{4}
  e^{\phi/2}\left(F^{MP}F^{N}_{\,\,\, P} -\frac{1}{4} F^2 G^{MN}
  \right)\tau \,h_{MN}\right], \label{h-d}\\ 
 S(V, \tau)&=& \int d^DX \sqrt{-G} \left[-\frac{1}{4} e^{\phi /2}
   F_{MN}\left( D^M V^N -D^N V^M\right) \tau\right], \\ 
S(V_2, V_2) &=& -\frac{1}{48} \int d^DX \sqrt{-G} \,e^{\phi}\, V_{[NP;
  M]}V^{[NP;M]},  \label{V_2-V_2}\\  
S(V,V_2) &=& -\frac{\kappa}{12} \gamma \int d^DX \sqrt{-G} \,
e^{\phi}\, V_{[NP;M]}F^{[MN}V^{P]}, \label{V-V_2}\eea 
where  $$F_{[MN}V_{P]}\equiv F_{MN}V_{P} + \, 2\,\, \mbox{cyclic
  perms}, \quad V_{[NP; M]}\equiv V_{NP; M} + \, 2\,\, \mbox{cyclic
  perms}.$$ 

We would like to remind the reader of the assumptions we have made to
derive (\ref{h-h}) and (\ref{V-V})-(\ref{V-V_2}) (and
(\ref{xi-xi})-(\ref{h-xi}) given in Appendix \ref{general-xi}): 
\begin{itemize}
 \item If the Kalb-Ramond field and the term $H_{MNP}H^{MNP}$ in
   (\ref{SB}) is not included, then the only assumption we made is
   that the background 
satisfies the EOMs (\ref{GMN-EOM})-(\ref{Y-EOM}).
\item  If the Kalb-Ramond field and the term $H_{MNP}H^{MNP}$ in
  (\ref{SB}) is instead included, we also assumed $D=6$ and the
  background gauge field $\mathcal{A}$ to lie in the Cartan
  subalgebra. 

\end{itemize}

We observe that if we want to focus on the $D$-dimensional EYM$\Lambda$
system we can restrict ourselves to the terms $S(h, h)$, $S(V,V)$ (for
$\gamma=0$), $S(h,V)$ and the $\xi$-dependent terms given in Appendix
\ref{general-xi}.  Instead, if we want to consider the
6D supergravity, we should put $\gamma=1$, $\mathcal{V}(\phi)=8\,g_1^2
\,e^{-\phi/2}/\kappa^4$ and also take into account the terms
(\ref{d-d})-(\ref{V-V_2}). 
Finally, we note that our results reduce to those of
Ref. \cite{RandjbarDaemi:2002pq} which studies a general
                                non-supersymmetric class of thick
                                brane models, once we take $T=0$,
                                $\gamma=0$ and we neglect the
                                fluctuations $V_{MN}$.

\subsection{Local symmetries}\label{local-symmetries}

As a consequence of the local symmetries of the complete model, the
linearized theory also possesses a number of local symmetries: 
\bea \delta h_{MN} &=& -\eta_{N;M} - \eta_{M;N}, \label{delta-h}\\
\delta V_M &=& -\eta^L F_{LM} - D_M \chi, \label{delta-V}\\ \delta
\tau &=& - \eta^M \partial_M \phi, \label{delta-tau}\\ 
\delta V_{MN}&=& 2 \gamma \kappa \,\chi F_{MN} +
\lambda_{N;M}-\lambda_{M;N}, \label{delta-V_2}\\ 
\delta \xi^M &=&
\eta^M(Y)-\zeta^{\alpha}\partial_{\alpha}Y^{M}.\label{delta-xi} 
\eea
Eqs. (\ref{delta-h}), (\ref{delta-V}) and (\ref{delta-tau}) represent
the effect of the local symmetries (descending from the
$D$-dimensional coordinate transformation invariance and gauge
symmetry) on the metric, the gauge field and the dilaton fluctuations
(see {\it e.g.} Ref. \cite{RandjbarDaemi:2002pq}). The bulk functions $\eta$ and
$\chi$ are the gauge functions associated with the $D$-dimensional
coordinate invariance and gauge symmetry. 

Eq. (\ref{delta-V_2}) represents instead a local symmetry acting on
$V_{MN}$, which descends from both the gauge symmetry and the
Kalb-Ramond symmetry\footnote{By Kalb-Ramond symmetry we mean the
  local invariance under $B_2\rightarrow B_2+d\Lambda$ of the action,
  where $\Lambda$ is a general 1-form.}. For this reason $\chi$ and
$\lambda_M$ are independent (bulk) gauge functions. Let us explicitly
check (\ref{delta-V_2}). To do so, it is enough to verify the
invariance of the 3-form (\ref{H-linear}) under (\ref{delta-V}) and
(\ref{delta-V_2}). We have 
\be \delta H^{(1)} = \frac{1}{\kappa} d\left(\delta V_2\right) +2
\gamma F\wedge \delta V =2\gamma d\left(\chi F\right) +2 \gamma
F\wedge \left( -\eta \cdot F -D\chi\right),\ee 
where we have used $d^2\lambda=0$ and $\eta\cdot F$ represents the
1-form with components $\eta^M F_{MN}$. Now, by using the 4D
Poincar\'e invariance  
of the background and $D=6$, which we always assume in the presence of
the Kalb-Ramond field, we have $F\wedge \left(\eta\cdot F\right)=0$
and $F\wedge \mathcal{A}=0$; also, by remembering that $\mathcal{A}$
is assumed to lie in the Cartan subalgebra, we have $dF=0$. These
equations are sufficient to conclude $\delta H^{(1)}=0$. 

Finally, Eq. (\ref{delta-xi}) represents the local transformation of
the perturbation of the brane position, descending from the
$D$-dimensional coordinate invariance and the 4D brane coordinate
transformation invariance (respectively the first and the second term
on the right hand side of (\ref{delta-xi})); the latter invariance is
associated to $\zeta^{\alpha}$ (a function of $x^{\alpha}$), which
represents another independent gauge function. 

\section{Perturbations in the Light Cone Static Gauge} \label{LCG}

Having derived the general bilinear action, we now have to choose a
gauge in order to study the physical spectrum. In this section we will
discuss our gauge choice and give the corresponding bilinear action.  

\subsection{Gauge fixing}\label{gauge-fixing}

We have two types of local symmetries: the bulk local symmetries
(which include the $D$-dimensional coordinate transformation invariance,
the gauge symmetry and the Kalb-Ramond symmetry) and the 4D coordinate
transformation invariance on the brane. Let us start with the first
group. 

A very convenient gauge choice for the bulk local symmetry is the
light cone gauge, as it ensures that the dynamics of sectors with
different spin decouple at the bilinear level\footnote{This has been
  observed in other studies, for example  
  \cite{Lutken:1987pc, Randjbar-Daemi:1984ap, Randjbar-Daemi:1984fs,
    RandjbarDaemi:2002pq}.}.
Another advantage of the light cone gauge is that it does not involve gauge
artifacts such as Faddeev-Popov ghosts, but contains only the physical 
spectrum \cite{Kaku, Lutken:1987pc, Randjbar-Daemi:1984ap}.  To define
this gauge, let us introduce $x^{(\pm)}\equiv \left(x^3\pm
  x^0\right)/\sqrt{2}$ and $A^{(\pm)}\equiv \left(A^3\pm
  A^0\right)/\sqrt{2}$, for a general vector $A^M$. Then the light
cone gauge is defined by  
\be V_{(-)}=0\,,\quad h_{(-)M}=0\,, \quad V_{(-)M}=0\,, \,\,\, \forall
\,M\,. \label{LCG2}\ee 
It can be proved that, after imposing (\ref{LCG2}), the $(+)$ components
of the different fields ({\it i.e.} $V_{(+)}$, $h_{(+)M}$ and $V_{(+)M}$) are not
independent, but can be expressed in terms of the other components by
means of constraint equations
\cite{Lutken:1987pc,Randjbar-Daemi:1984ap,RandjbarDaemi:2002pq}. We
therefore end up 
with the following independent bulk fields: $h_{ij}$, 
$h_{i \underline{m}}$, $V_i$, $V_{i \underline{m}}$, $h_{\underline{m}
  \underline{n}}$, $V_{\underline{m}}$, $V_{ij}$, $V_{\underline{m}
  \underline{n}}$ and $\tau$, where $i,j,...=1,2$. In particular the
$h_{(++)}$ field equation simply leads to the constraint 
\be h=0, \label{h++constraint}\ee
which brings a considerable amount of simplification. 

Concerning the 4D coordinate transformation invariance, we instead
impose the condition \cite{Sundrum:1998sj} 
\be \xi^{\mu}=0. \label{static}\ee
We will refer to (\ref{static}) as to the static gauge. We observe
that the light cone gauge and the static gauge are compatible because,
once we fix the light cone gauge by choosing $\eta_M$, $\chi$ and
$\lambda_M$ in a suitable way, we still have the freedom to perform
the local transformations generated by $\zeta^{\alpha}$. The static
gauge is also free from Faddeev-Popov unphysical ghosts
\cite{Sundrum:1998sj}. We observe that (\ref{static}) does not remove
completely the brane position fields $\xi^M$, but we are left with
their components along the extra dimensions
$\xi^{\underline{m}}\,$. We will refer to them as branons. Even if the
branons represent physical degrees of freedom, it can happen that they
can be consistently truncated {\it e.g.} by imposing an orbifold
symmetry, as in the RS models or in the 
conical-GGP compactification \cite{Parameswaran:2007cb}. In the
following we will confirm that the spin-0 fields $\xi^{\underline{m}}$ do not
have any mixing with the spin-2 and 
spin-1 sectors in the light cone gauge.

\subsection{Bilinear action in the light cone static gauge}
\label{S:finalbilinaction} 
Here we provide the bilinear action in the light cone static gauge,
that we have computed by imposing the gauge conditions (\ref{LCG2})
and (\ref{static}) on the general bilinear action and by using the
constraint equations for the $(+)$ components. In this section we assume
the form given in (\ref{static-background})-(\ref{background-H}) for
the background solution, and give the part of the action that
is independent of the branons.  Those involving the branons are given
in Appendix \ref{LCG-xi}.  

The results that are presented here reduce  to those for the
non-supersymmetric model present in\footnote{We do, however, correct
  some typos in that reference.}   
\cite{RandjbarDaemi:2002pq} once we take $T=0$, $\gamma=0$ and we
neglect the fluctuations $V_{MN}$; they also correctly reduce (for
$T=0$ and $\gamma=1$) to the results of \cite{Salvio:2007mb}, where
the linear perturbations of the sphere-monopole solution to the 6D
supersymmetric model are analyzed.

\subsubsection{Spin-2 action}
The spin-2 action $S^{(2)}$ only contains the field
$\tilde{h}_{ij}\equiv h_{ij}-\frac{1}{2} G_{ij}h_k^{\,\,\,k}$ and has
the following simple expression in terms of
$\tilde{h}_i^{\,\,\,j}=G^{jk}\tilde{h}_{ik}$: 
\be S^{(2)}(h,h)=-\frac{1}{4\kappa^2}\int d^DX \sqrt{-G}\, \partial_M
\tilde{h}_i^{\,\,\,j} \partial^M \tilde{h}_j^{\,\,\,i}. \label{spin-2}
\ee 
We observe that (\ref{spin-2}) has exactly the same form as in
\cite{RandjbarDaemi:2002pq} even if we have included the brane
terms. Therefore, the  brane sources do not explicitly contribute to
the spin-2 dynamics. We shall use (\ref{spin-2}) to derive the 4D
gravitational spectrum for the solutions described in Subsection \ref{EOM}.

\subsubsection{Spin-1 action} \label{spin-1-action}

The spin-1 action $S^{(1)}$ involves $h_{i \underline{m}}, V_i$ and
$V_{i \underline{m}}$. We have the following explicit expressions. 
\bea S^{(1)}(h,h) &=& \int d^DX \sqrt{-G} \left[-\frac{1}{2\kappa^2}
  \left(\partial_{\mu} h_{i\underline{m}} \partial^{\mu} h^{i
      \underline{m}} +\partial_{\rho} h_{i \underline{m}}
    \partial_{\rho} h^{i\underline{m}} + h_{i \underline{m} ; n}h^{i
      \underline{m} ; n}\right)     \right. \nonumber \\ 
&& -\frac{1}{4\kappa^2} h_{i m}h^{i m}\left( A'^2 + \frac{B'^2}{2}
\right)     -\frac{1}{4\kappa^2} h_{\rho i}
h_{\rho}^{\,\,\,i}\left(D_2A'B' -A'^2\right) \nonumber \\  
&& -\frac{1}{2} h_{i \underline{m}} h^{i}_{\,\,\, \underline{n}}
\left(\frac{1}{2} e^{\phi/2} F^{\underline{m}}_{ \quad
    \underline{l}}F^{\underline{nl}} +\frac{1}{4\kappa^2}
  \partial^{\underline{m}}\, \phi \,
  \partial^{\underline{n}}\,\phi\right)\nonumber \\ 
&&\left.+ \frac{1}{\kappa^2} A' h_{\rho}^{\,\,\,i} h_{m i}^{\quad ; m}
  -\frac{T}{4} \sqrt{g/G}\,\delta (X_c-Y_c) h_{\underline{m} \,i}\,
  h^{\underline{m} \,i}\right], \label{spin-1-hh}\eea 
where $'
\equiv
\partial_{\rho}$. The last term in (\ref{spin-1-hh}) is the brane
contribution. We have introduced the notation $X_c$ and $Y_c$ for the
internal components of the coordinate and the brane position
respectively, where the label $c$ stands for the codimension of the
brane.  The other non vanishing terms are the following.
\bea S^{(1)}(V,V) &=& \int d^DX \sqrt{-G}\,\, e^{\phi/2} \left[
  -\frac{1}{2} \left(\partial_{\mu} V_i \partial^{\mu} V^i +e^{-A}
    \partial_{\rho} V_i \partial_{\rho} V_i +D_mV_i D^m
    V^i\right)\right. 
\nonumber \\ && \left.-\frac{\kappa^2}{4} \gamma^2 e^{\phi/2} \left(
    F_{\underline{mn}}V_i\right)F^{\underline{mn}}V^i \right],  \\ 
S^{(1)}(h,V)&=&  \int d^DX \sqrt{-G}\,\, e^{\phi/2}
\left(-D_{\underline{m}}V_i h_{\underline{l}}^{\,\,\,i}
  F^{\underline{lm}}-\frac{1}{2} A' V_i h^{\underline{l}i}
  F_{\underline{l}\,\rho}\right), \\ 
S^{(1)}(V_2,V_2)&=& -\frac{1}{8} \int d^DX \sqrt{-G} e^{\phi}\left\{e^{-A}
\left[\partial_{\mu}V_{i \underline{m}} \partial^{\mu} V_{i}^{\,\,\,
    \underline{m}} 
\right.\right. \nonumber \\  &&\left.+G^{\underline{ml}}
G^{\underline{nh}} \left(\partial_{\underline{m}} V_{\underline{n
    }i}\partial_{\underline{l}}
  V_{\underline{h}i}-\partial_{\underline{m}} V_{\underline{n
    }i}\partial_{\underline{h}} V_{\underline{l}i}\right)\right]
\nonumber \\  
&& - e^{-4A -2\phi} \left(e^{\phi + 3A/2} \,V^{\underline{m}}_{\quad
    i}\right)_{;\underline{m}} \left(e^{\phi + 3A/2}\,
  V^{\underline{n}}_{\quad i}\right)_{;\underline{n}}  \nonumber \\
&&\left.  -2 e^{-2A} V_{\underline{m}i} 
\partial^{\underline{m}} \left[e^{-\phi-A/2 } \left( e^{\phi +3A/2}
    V^{\underline{n}}_{\quad i}
  \right)_{;\underline{n}}\right]\right\}, \\ 
S^{(1)}(V,V_2)&=& - \frac{\kappa}{2} \gamma \int d^DX \sqrt{-G}
e^{\phi} \left(-\frac{1}{2} A' V_{i m} V^i F_{\rho}^{\,\,\, m} +
  V_{\underline{n} i ; \underline{m}} F^{\underline{mn}} V^i
\right).\label{E:spin-1}\eea 
The term $S^{(1)}(h,V_2)$ vanishes as a consequence of $H_{MNP}=0$ (at
the background level). We observe that the term $S^{(1)}(V,V)$
reduces, as it should, to the corresponding action in
Ref. \cite{Parameswaran:2006db} in the case in which  $V_i$ is
orthogonal to the background gauge field.  Finally, we note that the
brane tension enters explicitly only in the term $S^{(1)}(h,h)$.

\subsubsection{Spin-0 action and singularities due to
  backreacting, fluctuating branes}\label{spin-0-action}

The last and most complicated part is the spin-0 action, which
involves\footnote{Note that $h_{\underline{mn}}$ and
  $h_{i}^{\,\,\,i}$ are not independent as Eq. (\ref{h++constraint})
  implies $h_{i}^{\,\,\,i}+h_{\underline{m}}^{\quad\underline{m}}=0$.}
$h_{\underline{mn}}$, $h_{i}^{\,\,\,i}$, $V_{\underline{m}}$, $\tau$,
$V_{ij}$, $V_{\underline{mn}}$ and $\xi^{\underline{m}}$. We observe
that, in the light cone static gauge, the fields $\xi^{\underline{m}}$
indeed only appear here. In other words they are in general
completely decoupled from the spin-2 and spin-1 fields.  Since it is
quite complicated, we give the explicit expression of the spin-0
action in Appendix \ref{LCG-xi}.  

Having completed the bilinear action, we should make some observations
regarding its consistency, in particular given the presence of
infinitesimally thin dynamical sources.
Indeed, as to be expected, if we include the gravitational
backreaction of the branes ($T \nrightarrow 0$) then there are
singular contributions to the dynamics of both the bulk
gravitational fluctuations and
the branons.

First, concerning the bulk
gravitational 
fluctuations, we encounter localized contributions to the mass terms in
both the spin-1 
(see Eq. (\ref{spin-1-hh})) and spin-0 (see Eq. (\ref{spin-0-hh}))
sectors.  These 
contributions involve the behaviour of background and perturbed
fields at the background positions of the branes.  They are
well-defined in the codimension one RS scenario, where the metric is
well-defined everywhere 
including at the brane 
positions (although its derivatives are not).  They do not
appear to be 
well-defined in the codimension two (or higher) case, where the
internal metric is 
actually singular at the brane positions due to the backreaction of
the branes. However, as we shall see in subsequent sections, these
terms do not obstruct our derivation of the 4D particle spectra
arising from bulk modes in 6D.

Meanwhile, the linearized dynamics for the branons of a backreacting
brane would be more
problematic. 
For example, in 
(\ref{spin-0-xi-xi}), since the action is evaluated at the
background position of the branes, the kinetic
term for the branons is not well-defined in the codimension two case,
because of the conical defect in $G_{mn}$.
Such a singularity was discussed in \cite{Sundrum:1998ns}, where it
was argued that within the domain of validity of the effective field
theory, the curvature singularity could be discarded.  Moreover, the
mass term for the branons takes the form\footnote{$\delta(0)$
  singularities due to the  
  localization of fields on a boundary have been discussed in
  a different context (5D SYM theory on $S^1/Z_2$) in \cite{delta0}.} of a
$\delta(0)$.  These singularities
are not present in the RS model, as there the 
branons are projected out with an
orbifolding\footnote{Indeed, in the
  RS literature, the radion 
has been studied in depth \cite{radion}, but the branons are absent.
Although the radion can also 
  be seen as a brane bending in the case of RS, since the branes are
  at the boundaries of the internal space,
  one should not confuse the radion with the branon.  The radion 
  is a deformation of the bulk metric, whereas the branon is a deformation
  of the brane itself within the bulk manifold. As a check of our
  formalism, we will find the radion mode in Subsection
  \ref{0spin-0}.}.  
Indeed we should reiterate here that we
apply our analysis to codimension one branes only on orbifold fixed
points (to avoid the appearance of Gibbons-Hawking boundary terms),
and so without branon degrees of freedom.  

In order to perform a complete analysis of the spin-0 action in
codimension two (or higher) models, taking into account both the
backreaction of the brane and its dynamical fluctuations, it seems
necessary to resolve the thin structure of the brane.  Otherwise we
can assume a brane tension much smaller than the 6D fundamental scale,
so that its backreaction is negligible.  Or we can assume a high brane
tension so that the brane is very heavy and rigid and does not
oscillate.  Or else we can assume an additional orbifold symmetry
under which the branons are projected out  -- an example of such a
symmetry has been provided in Ref. \cite{Parameswaran:2007cb} and is
discussed in\footnote{By using the explicit expression  
  for the mixing terms between branons and bulk fields given in
  Appendix \ref{LCG-xi}, it is easy to confirm that symmetry
  consistently truncates the branons.} Appendix \ref{fullspin0}.  
In these cases, we can avoid the singular branon action.

\section{6D (and 5D) Braneworlds} \label{Applications}
In the second part of this paper, we apply the results of the previous sections to derive the 4D particle spectra in specific
setups.  Our main interest is in the warped (and unwarped)
axi-symmetric braneworld compactifications of 6D supergravity, but
along the way we shall also discuss the rugbyball compactifications in
the non-supersymmetric 6D EYM$\Lambda$ theory, as well as the 5D
Randall-Sundrum models.  We discuss in order the spin-2, spin-1 and
spin-0 fluctuations.  

\subsection{Gravitational fluctuations} \label{grav-fluct}
The simplest application of our results is the analysis of the spin-2
sector. As we have discussed, this sector completely decouples from
the rest.  The $\tilde{h}_i^{\,\,j}$ fields contain only the maximal
helicity components of a spin-2 multipet; one should look for the
remaining components in the spin-1 and spin-0  actions.  However, by
virtue of 4D Poincar\'e invariance, the lower helicity components must
have the same spectrum \cite{RandjbarDaemi:2002pq}. We can therefore
focus on Eq. (\ref{spin-2}) to study the spin-2 fluctuations.  

In order to analyze this sector we deduce the EOMs from Eq. (\ref{spin-2}):
\be \partial_M \left( \sqrt{-G} \,\partial^M \tilde{h}_{i}^{\,\,
    j}\right)=0 \qquad  \forall \,i, j \,.\label{spin-2-EOM}\ee  
In deriving this equation we have required as usual that the boundary
terms which emerge in the integration by parts vanish, that is  
\be \int d^DX \partial_M\left(\sqrt{-G} \,\delta \tilde{h}_i^{\,\, j}
  \partial^{M}\tilde{h}_i^{\,\,j}\right)=0,\label{spin-2-BC} \ee 
where $\delta \tilde{h}_i^{\,\, j}$ is the variation of the field
$\tilde{h}_i^{\,\, j}$, which is performed to apply the minimal action
principle.   Since we assume standard boundary conditions on the 4D
boundary, (\ref{spin-2-BC}) reduces to
\cite{Nicolai:1984jg,Parameswaran:2006db} 
\be \int d^{D_2+1}X \partial_{\underline{m}}\left(\sqrt{-G} \,\delta
  \tilde{h}_i^{\,\, j}
  \partial^{\underline{m}}\tilde{h}_i^{\,\,j}\right)=0.\label{spin-2-BC2}\ee 
We now perform a KK decomposition of the fields as follows:
\be \tilde{h}_i^{\,\, j}(X)= \sum_{\bf k} \tilde{h}^{({\bf k})j}_i(x)
f_{\bf k}(\rho , y),\ee
where ${\bf k}$ represents a collective KK number. 
By taking $\tilde{h}^{({\bf k})j}_i(x)$ to be an eigenfunction of
$\eta^{\mu \nu}\partial_{\mu}\partial_{\nu}$, that is $\eta^{\mu
  \nu}\partial_{\mu}\partial_{\nu}\tilde{h}^{({\bf
    k})j}_i(x)=M_{\bf k}^2\, \tilde{h}^{({\bf k})j}_i(x)$, the EOMs
(\ref{spin-2-EOM}) become  
\be -\frac{1}{\sqrt{-G}}e^A \partial_{\underline{m}}\left(\sqrt{-G}
  \partial^{\underline{m}}f_{\bf k}\right) = M_{\bf k}^2 f_{\bf k}
\label{spin-2-EOM2}\ee 
 and the corresponding boundary conditions (\ref{spin-2-BC2}) read (we
 recall that  
$\delta \tilde{h}_i^{\,\, j}$  and $\tilde{h}_i^{\,\, j}$ are
independent fields) 
\be \int d^{D_2+1}X \partial_{\underline{m}}\left(\sqrt{-G} \,f_{{\bf
      k}'}\partial^{\underline{m}}f_{\bf k}\right)=0. \qquad \forall
\, {\bf k}, {\bf k}'\,. \label{HC}\ee 
Condition (\ref{HC}) ensures that the operator acting on $f_{\bf k}$ in the left hand side of (\ref{spin-2-EOM2}) is a Hermitian operator \cite{Nicolai:1984jg,Parameswaran:2006db}; we will therefore refer to (\ref{HC}) as the hermiticity condition (HC).
In addition to the HC we will also require the wave functions $f_{\bf k}$ to be normalizable, that is $$\int d^{D_2 +1}X \sqrt{-G} e^{-A}f_{\bf k}^2 < \infty.$$ This  normalizability condition (NC) is equivalent to the finiteness of the kinetic energy of the modes $\tilde{h}^{({\bf k})j}_i(x)$.  
We observe that there is always a constant massless ($M^2_{\bf k}=0$)
solution to (\ref{spin-2-EOM2}), satisfying the HC (\ref{HC}). This
solution corresponds to a 4D graviton provided that the NC is satisfied,
that is $\int d^{D_2 +1}X \sqrt{-G} e^{-A}< \infty$. 

\subsubsection{Randall-Sundrum}

In the special case $D=5$, and therefore in particular for the RS
background (\ref{RS-solution}), the EOM (\ref{spin-2-EOM2}) has the
form 
\be -e^{-A}\partial_{\rho}\left(e^{2A} \partial_{\rho}f_{\bf k}\right)
= M_{\bf k}^2  f_{\bf k} \,. \label{RS-spin-2}\ee
Here we do not want to analyze the latter equation as this has been
done in the original RS works, but we observe, as a check of our
spin-2 action, that (\ref{RS-spin-2}) has exactly the same form as in
\cite{{RSb}}. 

\subsubsection{6D Brane Worlds} \label{6DBraneWorlds}
\begin{figure}
\centering
\begin{tabular}{cc}
\includegraphics[scale=0.80]{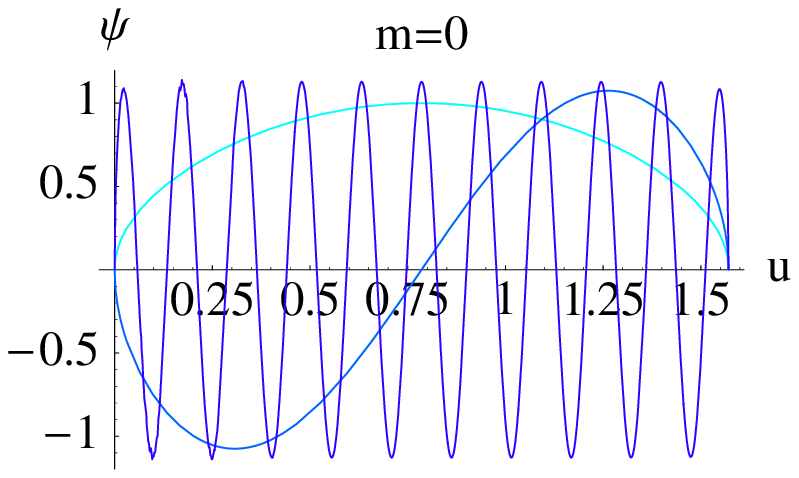} &
\includegraphics[scale=0.80]{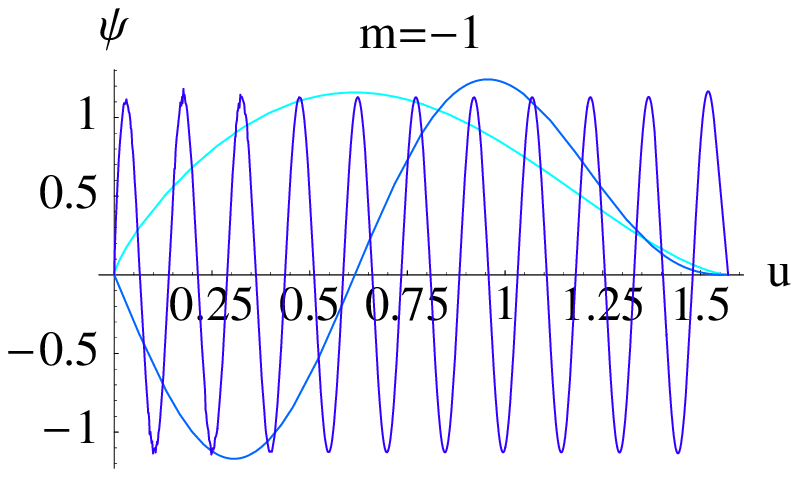}  
\end{tabular}
\caption{\footnotesize Graviton Wave Function Profiles: ${\bf n} = 0,
  1, 20$ modes plotted for angular momentum numbers 
${\bf m} =-1, 0$. The parameters are chosen to be $(r_0, \omega,
\overline{\omega}) = (1, 1/4, 1)$, corresponding to a single negative 
tension brane at $u = 0$. Also the normalization constant is set such that
$\int 
du |\psi|^2 = 1$. The number of
intersections with the u-axis equals $n$, according to quantum
mechanics. Notice that the $({\bf m}, {\bf n}) = (0, 0)$ 
mode is massless.}
\label{fig:wavefns}
\end{figure}

We now move to the conical-GGP solutions to 6D supergravity given in
Eqs. (\ref{GGPsolution})-(\ref{GGPsolution2}). Since 
our internal space is topologically $S^2$, we require $\tilde{h}_i^{\, j}$
to be periodic functions of $\varphi$:
\be \tilde{h}_i^{\,\, j}(X)= \sum_{{\bf n},{\bf m}} \tilde{h}_{i\,{\bf n
    m}}^{\,\, j}(x)f_{{\bf n  m}}(\rho) e^{i{\bf m}\varphi},
\label{spin-2-exp}\ee 
where ${\bf m}$ is a generic integer and ${\bf n}$ is an extra KK
number that emerges as we have a number of compact dimensions greater
than one.  Also we observe that Eq. (\ref{spin-2-EOM2}) with the HC
and NC is formally identical\footnote{In \cite{Parameswaran:2006db}
  there is the extra parameter $N_V$, which is equal to zero here. To
  check that the two problems are identical it is useful to remember 
$A=\phi/2$, which is true for the conical-GGP solutions. Also, take
care that
$\phi$ in reference \cite{Parameswaran:2006db} is half $\phi$ here.}
to the corresponding problem for 4D gauge fields addressed in
Ref. \cite{Parameswaran:2006db}. Therefore, here we only give the
result. The wave functions can be expressed in a more compact way by
introducing  
\be \psi \equiv e^{(3A+B)/4} f, \label{fpsi}\ee
where we have suppressed ${\bf n}$ and ${\bf m}$. The explicit
expression for $\psi$ is  
\be \psi \propto  z^{\epsilon}(1-z)^{\beta}F(a,b,c,z), \label{psi1} 
\ee
where $z\equiv\cos^2\left(u/r_0\right)$, $F$ is Gauss's hypergeometric
function and 
\bea \epsilon &\equiv& \frac{1}{4}\left(1+2|{\bf
    m}|\overline{\omega}\right), \,\, 
\beta\equiv \frac{1}{4}\left(1+2{\bf m}\omega\right),
\,\, c\equiv 1+|{\bf m}|\overline{\omega}, \nonumber \\
a&\equiv&\frac{1}{2}+\frac{{\bf m}}{2}\omega+\frac{|{\bf
    m}|}{2}\overline{\omega} 
+\frac{1}{2}\sqrt{r_0^2M^2+1+{\bf m}^2\left(\omega-\overline{\omega}\right)^2},
\nonumber\\
b &\equiv&\frac{1}{2}+\frac{{\bf m}}{2}\omega+\frac{|{\bf
    m}|}{2}\overline{\omega} 
-\frac{1}{2}\sqrt{r_0^2M^2+1+{\bf m}^2\left(\omega-\overline{\omega}\right)^2},
\label{gbetaabcV}\eea
with
\be \omega\equiv(1-\delta/2\pi)^{-1}, \qquad
\overline{\omega}\equiv(1-\overline{\delta}/2\pi)^{-1}.\ee
Moreover the explicit form of the mass spectrum is given by
\be M^2= \frac{4}{r_0^2} \left[{\bf n}({\bf n}+1) + \left(\frac{1}{2}
    + {\bf n}\right)|{\bf m}| \left(\omega +\overline{\omega}\right)
  +{\bf m}^2 \omega \overline{\omega}\right] \geq 0,
\label{spin-2-masses}\ee  
where ${\bf n}=0,1,2,3,...$ \cite{Parameswaran:2006db}.
So we have obtained the exact and complete spectrum (wave
functions and masses) for the spin-2 fluctuations of the conical-GGP
solutions. We observe that Eq. (\ref{spin-2-masses}) tells us there is a
massless normalizable solution (for ${\bf n}={\bf m}=0$), which
corresponds to the 4D graviton. This solution is separated from the
first KK excitation by a finite mass gap, which is of order $1/r_0$
(if $\omega \sim \overline{\omega}\sim 1$). 
We plot some representative wave function profiles
in Figure \ref{fig:wavefns}. As discussed in
\cite{Parameswaran:2006db} the asymptotic behaviour close to the
branes is universal for each KK tower, and it does not appear possible
to separate the infinite number of heavy 
modes from the light ones by using their respective wave function profiles.

Here we also observe that Eq. (\ref{spin-2-EOM}) is independent of
$\gamma$ and the dilaton potential $\mathcal{V}$. This implies that
the spin-2 spectrum of the non supersymmetric and supersymmetric
models are the same (provided the backgrounds are the same). Indeed,
the rugbyball configuration (that is $\omega=\overline{\omega}$) leads
to the same spin-2 spectrum in the EYM$\Lambda$ model and in the 6D
supergravity.  

Finally, as a check, we can consider the $S^2$ limit
($\omega,\overline{\omega} \rightarrow 1$), whose mass spectrum is
well-known.  Our spectrum (\ref{spin-2-masses}) reduces to
\be \frac{r_0^2}{4}M^2=l(l+1),\quad
\mbox{multiplicity}=2l+1 \, ,\label{sphere-spin-2}\ee
where $l=0,1,2,3,...$. Since $r_0/2$ represents the radius of $S^2$
in the sphere limit, this is exactly the result that one finds by
using the spherical harmonic expansion \cite{sphere,Salam:1984cj} from
the beginning. 

\subsection{Vector fluctuations}\label{vector fluctuation}

Here we analyze the vector fluctuations, in particular their
wavefunction expansions and mass spectra.  In the following subsection
we shall study the implications of these results for the structure of
the 4D gauge group in the 6D models of interest. The physical 4D vector field
spectrum can be extracted from the spin-1 action given in Subsection
\ref{spin-1-action}. However, some of the perturbations in that action
are simply the helicity-($\pm 1$) components of massive gravitons and
therefore should not be interpreted as independent vector fields.  

\subsubsection{Randall-Sundrum}

To illustrate the previous point we first notice that our spin-1 action leads
to the well-known result that there are no physical 4D vector fields
in the RS model (unless one introduces bulk gauge fields). Indeed, in
that case the only field appearing in the spin-1 action is $h_{
  i\rho}$, whose action is simply 
\be S^{(1)}(h,h) \propto \int d^5X \sqrt{-G}
\left(\partial_{\mu}h_{i\rho} \partial^{\mu}h^{i}_{\,\,
    \rho}+\partial_{\rho}h_{i\rho} \partial_{\rho}h^{i}_{\,\,
    \rho}-\frac{A'^2}{2}h_{i\rho} h^{i}_{\,\, \rho}-\frac{3}{2}
  A''h_{i\rho} h^{i}_{\,\, \rho}\right),\ee 
where we used the property
\be A'' +T\kappa^2 \delta( \rho -Y)/3=0 \label{RSproperty} \ee
that follows from the form of the warp factor,
Eq. (\ref{RS-solution}), in the RS 
model. Therefore, once (\ref{RSproperty}) is used the problem assumes
the same form as in Ref. \cite{RandjbarDaemi:2002pq}, where it is
shown that the 4D spectrum from $h_{i \rho}$ exactly reproduces the
graviton one with the zero mode removed. By counting the degrees of
freedom, it follows that there are no physical vector fields. 

\subsubsection{6D Brane Worlds}\label{6D-brane-1}

Let us begin by considering what we might expect from the 
symmetries of the problem -- with some benefit of hindsight from the
authors.  In the limit where the brane tensions go to zero, the smooth
sphere-monopole compactification is recovered.  In this case, standard
KK theory tells us that there are three massless KK gauge
bosons\footnote{There may be additional massless gauge fields
  descending from any unbroken higher dimensional gauge symmetries.},
which manifest the $SU(2)$ isometries of the sphere in the 
4D theory \cite{sphere}.  Clearly, any branes break the spherical symmetry 
in the internal dimensions.  For the solutions of present interest
an axial isometry survives, and therefore we can expect the 4D
theory to enjoy a $U(1)$ KK gauge symmetry\footnote{For the analysis
  of a very similar model, in which the $SU(2)$ KK gauge symmetry of
  the sphere is broken down to $U(1)$ by {\it smooth} axisymmetric
  deformations, see Ref. \cite{RandjbarDaemi:2006gf}. }.

At the same time, for the case of an unwarped
``saddle-sphere'' with the special deficit angles (for $\alpha=1$ we
recover the sphere) 
\be
\delta = -2\pi, -4\pi, \dots \qquad \mbox{or} \qquad \alpha =
\frac{1}{\omega} = 2,3,\dots 
\label{special} 
\ee
the metric (\ref{rugbyball-metric}) -- defined everywhere but at the
branes -- has three 
single-valued Killing vectors, which obey the Lie algebra of $SU(2)$:
\be K^+ = e^{i \alpha \varphi} \left(\frac{\partial}{\partial \theta}
  +  i \cot
\theta \frac{1}{\alpha}\frac{\partial}{\partial
  \varphi} \right), \,\, K^-= e^{-i\alpha\varphi}
\left(-\frac{\partial}{\partial 
  \theta} + i \cot \theta \frac{1}{\alpha}\frac{\partial}{\partial
  \varphi}\right), \,\, K^0= 
-\frac{i}{\alpha}\frac{\partial}{\partial \varphi}. \label{kvf}
\ee
Only one of these Killing vectors, $K^0$, implies a genuine continuous isometry,
since $K^{\pm}$ cannot be globally integrated to an isometry\footnote{To
  avoid the need of differential geometric results for singular
spaces, we can consider removing the brane singularities and
taking instead a smooth non-compact manifold, for which $0 <
\theta < \pi$. 
Killing vector fields are the
generators of the {\it infinitesimal isometries} of a manifold,
whereas an isometry is a global aspect of the geometry.  
Whilst for smooth compact manifolds the Killing vectors are always
globally integrable to an isometry,  
for non-compact manifolds this may not always be the case.}.
In other words, we have an {\it infinitesimal} $SU(2)$ isometry for
the special saddle-spheres, compared to a genuine $SU(2)$ isometry for
the sphere.  As we will show, this turns out to be sufficient to
ensure three massless 4D vectors amongst the KK spectra.  From the point
of view of the full 4D theory, however, we will argue that these
massless fields arise accidentally and that their masslessness is not
protected by any symmetry. 

\paragraph{Rugbyball Harmonics}

Let us now see how the above story plays out in detail.  Our focus
shall then be on the unwarped rugbyballs and saddle-spheres,
Eq. (\ref{rugbyball-metric}), and 
indeed all previous results have indicated that warping does not
lead to any qualitative changes in the physics (see the spin-2 results
in the present paper, as well as
Refs.\cite{Parameswaran:2006db}-\cite{Lee:2006ge}).  
We shall thus proceed by finding a set of ``rugbyball harmonics'' and their
mass spectra, in analogy to the spherical harmonics (and, more
generally, the so-called Wigner functions) used in the smooth sphere
compactification \cite{sphere}.  

We first observe that the vector field fluctuations $V_i$, which are
orthogonal to the gauge field background ($V_i
F_{\underline{mn}}=0$), do not mix with the other perturbations $h_{i
  \underline{m}}$ and $V_{i \underline{m}}$. These fluctuations have
been already studied in Ref. \cite{Parameswaran:2006db} where the
complete KK towers are provided and it is shown that there are as many
4D gauge fields as fluctuations $V_i$ with vanishing monopole charge
($V_i \times F_{\underline{mn}}=0$), as expected from group
theory. Here we therefore consider only the case when $V_i$ is
parallel to the background monopole. 

Now, from Subsection \ref{spin-1-action} it follows that the spin-1
action for unwarped solutions has the following form. 
\bea S^{(1)}(h,h)&=& -\frac{1}{2\kappa^2}\int d^6X \sqrt{-G}
\left(\partial_{\mu}h_{mi}\partial^{\mu}h^{mi} +h_{mi;n}h^{mi;n}
  +\frac{R}{2}h_{mi} h^{mi}\right), \nonumber \\ 
S^{(1)}(V,V) &=& -\frac{1}{2} \int d^6X
\sqrt{-G}\left[\partial_{\mu}V_i\partial^{\mu} V^i+ \partial_{m}V_i
  \partial^{m}V^i + \frac{\kappa^2}{2}\gamma^2 F^2 V_i V^i\right],
\nonumber \\ S^{(1)}(V_2,V_2) &=& -\frac{1}{8}\int d^6X \sqrt{-G}
\left(\partial_{\mu}V_{mi}\partial^{\mu}V^{mi} +V_{mi;n}V^{mi;n}
  +\frac{R}{2}V_{mi} V^{mi}\right),\nonumber \\ 
S^{(1)}(h,V)\hspace{-0.25cm} &+&\hspace{-0.25cm} S^{(1)}(V,V_2)=  \int
d^6X \sqrt{-G}\left(-\partial_m V_i h_{ni} F^{nm}-\frac{\kappa}{2}
  \gamma \partial_m V^i V_{ni} F^{nm}\right),
\label{rugbyball-spin-1}\eea 
where we have  used $V_{i;m}= \partial_m V_i$
because the background solution is unwarped and $V_i$ is uncharged under the
background monopole. Also $m,n,...$ here run over $\theta$ and $\varphi$.
To derive the last term in $S^{(1)}(h,h)$ we have used the Einstein equations: 
\be \frac{2}{\kappa^2}\sqrt{-G}R_{mn} = \sqrt{-G}F_{ml} F_{n}^{\,\,
  \,l} +G_{mn} T\delta(X_2 - Y_2),\label{spin-1-system}\ee 
which allow us to rewrite the brane contribution in the last term of
(\ref{spin-1-hh}) as a combination of the Ricci tensor and the field
strength. Also we have used that in two dimensions $R_{mn}=G_{mn} R/2$
and the Maxwell equations $F^{mn}_{\,\,\quad ; m}=0$. The EOMs descending
from (\ref{rugbyball-spin-1}) are the following. 
\bea &&\left(\partial^2 +D^2 -\frac{R}{2} \right)h_{mi} -
\kappa^2F_m^{\,\,\,\,l}\partial_lV_i =0, \label{h1-EOM} \\ 
&&\left(\partial^2 +D^2 -\frac{R}{2} \right)V_{mi} - 2\kappa \gamma
F_m^{\,\,\,\,l}\partial_lV_i =0,\label{Vm2-EOM} \\ 
&& \left(\partial^2 +D^2 -\frac{\kappa^2}{2}\gamma^2 F^2 \right)V_i +
F^{nm}h_{ni;m} +\frac{\kappa}{2}\gamma  F^{nm}V_{ni;m} =0,
\label{V-EOM}  
 \eea
where $\partial^2\equiv \partial_{\mu}\partial^{\mu}$ and $D^2\equiv
D_mD^m$.  As for the spin-2 case above, the EOMs come with a set of boundary
conditions, which we refer to as Hermiticity Conditions (HCs)
\cite{Nicolai:1984jg,Parameswaran:2006db}:  
\bea &&\int d^6X \sqrt{-G} \left(\delta h_{mi}  h^{mi;n}\right)_{;n}
=0, \quad \int d^6X \sqrt{-G} \left(\delta V_{mi} V^{mi;n}\right)_{;n}
=0, \label{hm-HC}\\ &&\int d^6X \sqrt{-G} \left(\delta V^i h_{li}
  F^{lm} \right)_{;m} =0, \quad \int d^6X \sqrt{-G} \left(\delta V^i
  V_{li} F^{lm} \right)_{;m} =0, \label{hV-HC}\\&& \int d^6X \sqrt{-G}
\left(\delta V_i \partial^m V^i\right)_{;m}=0. \label{V-HC}\eea 
We will additionally impose the usual Normalizability Conditions (NCs).

We can immediately observe that there is a simple solution to
Eqs. (\ref{h1-EOM})-(\ref{V-EOM}), with $h_{\theta i}=h_{\varphi i}=0$,
$V_{\theta i}=V_{\varphi i}=0$ and $V_i$ independent of the extra
dimensions.  Its squared mass is given by 
\be M^2 = \frac{\kappa^2}{2}\gamma^2 F^2= \frac{8}{r_0^2}\gamma^2,
\label{6DU(1)}\ee 
where we explicitly used the rugbyball solution. We see that the
monopole $U(1)$ is a gauge symmetry in the EYM$\Lambda$ model
($\gamma =0$), whereas it is broken in 6D supergravity ($\gamma=1$), like
for the smooth  
sphere-monopole solution \cite{seifanomaly}. 

We now want to find the general solution to
Eqs. (\ref{h1-EOM})-(\ref{V-EOM}) subject to the HCs and NCs. 
System (\ref{h1-EOM})-(\ref{V-EOM}) is a rather complicated set of
coupled differential equations, but the case
at hand can be elegantly solved by using the harmonic expansion of the
scalar Laplacian; let us now describe this technique.   
It is easy to solve the eigenvalue problem for $-D^2$ acting on the
2D scalars (in fact, for $A=0$ and applying the diagonal HC and the NC,
the system is identical to that for the helicity-2 field above).  The
eigenfunctions are then given by (\ref{spin-2-exp}-\ref{gbetaabcV})
with $\omega=\overline{\omega}$, and the eigenvalues can be written as:
\be \mu^2_{{\bf n \, m}}\equiv \frac{4}{r_0^2}\left({\bf n}+|{\bf m}| \omega\right)\left({\bf n}+|{\bf m}| \omega+1\right)\geq 0, \label{mu}\ee 
 where ${\bf n}=0,1,2,3,...$ and ${\bf m}$ run over all the integers.
 This is the generalization to the rugbyball of the scalar spherical
 harmonics.  They form a complete basis for the 2D scalar fields $V_i$. 

 We next proceed by determining a complete basis for the 2D vector
 fluctuations $\{h_{\theta i},h_{\varphi i}\}$ and
 $\{V_{\theta i},V_{\varphi i}\}$. We focus only on
 $\{h_{\theta i},h_{\varphi i}\}$ 
as the analysis
 for $\{V_{\theta i},V_{\varphi i }\}$ is identical. A way to determine such
 a basis is to look at the eigenvalue problem for the operator $-D^2 +
 R/2$ appearing in Eq. (\ref{h1-EOM}), because the
 diagonal HC (\ref{hm-HC})
 guarantees that this operator is Hermitian over the space of
 functions where $\{h_{\theta i},h_{\varphi i}\}$ lives, and therefore has a
 complete basis of eigenfunctions. 
Again, this system is easy to solve, this time using the
results\footnote{In Ref. \cite{Parameswaran:2007cb} a more general 
   problem has been solved, which reduces to the present one in the
   unwarped case $A=0$.} of Ref. \cite{Parameswaran:2007cb}. We
 therefore just summarise the results.
The eigenvalue problem for $-D^2 + R/2$ on 2D
 vectors generically mixes the $h_{\theta i}$ and $h_{\varphi i}$
 components, but reduces to a diagonal form, at least in the
 rugbyball case, by introducing\footnote{The $\pm$ appearing in
   (\ref{pm}) and throughout this section should not be confused  
with the $(\pm)$ used to defined the light-cone gauge in Subsection
\ref{gauge-fixing}, for this reason the latter are written inside
brackets.} 
\be h_{\pm i}\equiv \frac{1}{\sqrt{r_0}} \left(e^{B/4}h_{\theta i} \pm
  i e^{-B/4} h_{\varphi i}\right).\label{pm}\ee 
Eq. (\ref{pm}) defines a new basis for tensors on the
2D internal space, and we remind the reader that for the rugbyball
$e^B = \alpha^2 \sin^2\theta$.  
The squared mass problem can then be transformed into a pair of decoupled
Schr\"odinger equations, which can be solved. Note that the
singularities of the spin-1 
action discussed in Subsection \ref{spin-0-action} appear in the
Schr\"odinger problems as two singular points in the effective
potentials (one for each brane), which do not obstruct the
determination of the spectrum
\cite{Parameswaran:2006db,Parameswaran:2007cb}. The $h_{\pm i}$ fields
can be KK expanded as follows 
\be h_{\pm i}(X)= \sum_{{\bf n},{\bf m}}h_{\pm i \,\,{\bf n \, m}}(x)
f^{\pm}_{{\bf n \, m}}(\theta) e^{i{\bf m}\varphi}, \label{pmkk}\ee
and, in the case 
\be {\bf m}  = 0 \quad \mbox{or} \quad |{\bf m}|  \geq 1/\omega,
\label{validity}\ee 
 both the KK tower associated with $f^{+}$ and $f^{-}$ turn out to be
 exactly that in (\ref{mu}), where ${\bf n}=0,1,2,3,...$ and 
${\bf m}$ run over all the integers, but with the constraint  $\{{\bf
  n},{\bf m}\}\neq \{0,0\}$.  Condition
(\ref{validity}) is satisfied by every $|{\bf 
  m}|$ for  non-negative tensions and it is satisfied by some (but not
all) $|{\bf m}|$ for negative tensions. This, however, will be enough
to show that when the tensions assume the values in (\ref{special}),
the KK spectra include extra massless spin-1 fields. In the following
we denote (\ref{validity}) with $0 \not< |{\bf m}|\omega \not< 1$. 

So we have found that, for modes satisfying (\ref{validity}), the spectrum 
of $-D^2 + R/2$ on 2D vectors is made up of two identical copies of the
spectrum of $-D^2$ on 2D scalars 
but with zero mode removed. This suggests that we may be able to
express the eigenfunctions of $-D^2 + R/2$ on 2D vectors in terms of
eigenfunctions of $-D^2$ on 2D scalars. Indeed, if we consider a solution $V$
to the eigenvalue problem of  $-D^2$ with eigenvalue $\mu^2$, then it
is easy to show that $\partial_m V$ is an eigenfunction of $-D^2 +
R/2$ with the same eigenvalue. In the case (\ref{validity}), this
implies that we can write   
\be \partial_{\pm} \left(f_{{\bf n \, m}}(\theta) e^{i{\bf m}\varphi}
\right) = c_{{\bf n \, m}} f^{\pm}_{{\bf n \, m}}(\theta) e^{i{\bf
    m}\varphi}, \label{derivrel} \ee 
where we have used the basis defined in Eq.(\ref{pm}) for
$\partial_{\pm}$, 
and moreover $c_{{\bf n \, m}}$ are normalization constants which,
having chosen a convenient normalization for 
the wave functions,  can be fixed to be
$c_{{\bf n \, m}}= \mu_{{\bf n \, m}}/\sqrt{2}$.  This is the analogue
of the derivative relation that exists between Wigner functions for
fields of different spin on the sphere (see eq. (3.17) of 
Ref. \cite{sphere}).  So, remarkably, we can construct the complete
harmonic expansion for\footnote{It is easy to show that all we have
  stated about the harmonic expansion for $h_{mi}$ holds for $V_{mi}$ as
  well.} $V_i$, $h_{\pm i}$ and $V_{\pm i}$ by using the solution to the
eigenvalue problem for the scalar Laplacian.  
Moreover, it is easy to check that, having applied the diagonal HCs to
derive the complete basis for $h_{\pm i}$ and $V_{\pm i}$, the mixed
HCs (\ref{hV-HC}) are automatically satisfied. 

Having derived the harmonic expansions one more observation is
necessary.  It turns out that the factor $F_+^{\,\,\,+}$ which appears
in the mixing terms $S^{(1)}(h,V)$ and $S^{(1)}(V_2,V)$ is constant
for the rugbyball ($F_+^{\,\,\,+}=2\sqrt2 i /(r_0 \kappa)$).  Putting
everything together, we are then able to transform the differential
eigenvalue problem for the squared mass operator into an algebraic
problem that can be solved. In particular, after integrating out the
extra dimensions, Action (\ref{rugbyball-spin-1}) assumes the
following form in terms of the KK modes. 
\bea &&S^{(1)}(h,h)+S^{(1)}(V_2,V_2)  +
S^{(1)}(h,V) + S^{(1)}(V,V_2) +
S^{(1)}(V,V)\nonumber \\ &=&\int d^4 x \sum_{{\bf n},{\bf
    m}}\left\{\frac{1}{2\kappa^2}\left(h_{+i\,\,{\bf n \, m}}\right)^*
  \left(\partial^2
    -\mu^2_{{\bf n \, m}}\right)h_{+i\,\,{\bf n \,
      m}}+\frac{1}{2\kappa^2}\left(h_{-i\,\,{\bf n \, m}}\right)^*
  \left(\partial^2 
    -\mu^2_{{\bf n \, m}}\right)h_{-i\,\,{\bf n \, m}}\right. \nonumber \\  
&&+\frac{1}{8}\left(V_{+i\,\,{\bf n \, m}}\right)^* \left(\partial^2
  -\mu^2_{{\bf n \, m}}\right)V_{+i\,\,{\bf n \,
    m}}+\frac{1}{8}\left(V_{-i\,\,{\bf n \, m}}\right)^* \left(\partial^2
  -\mu^2_{{\bf n \, m}}\right)V_{-i\,\,{\bf n \, m}}\nonumber \\ 
&&-\frac{2\mu_{{\bf n \, m}}i}{r_0\kappa}\left[\left(h_{+i\,\,{\bf n \,
        m}}\right)^* V_{i\,\,{\bf n \, m}}-\left(h_{-i\,\,{\bf n \,
        m}}\right)^*V_{i\,\,{\bf n \, m}}\right.\nonumber \\
&&\left.+\frac{\kappa}{2}\gamma 
  \left(\left(V_{+i\,\,{\bf n \, m}}\right)^*V_{i\,\,{\bf n \,
        m}}-\left(V_{-i\,\,{\bf n \, m}}\right)^*V_{i\,\,{\bf n \,
        m}}\right)\right]\nonumber \\  
&&\left.+\frac{1}{2} \left(V_{i\,\,{\bf n \, m}}\right)^*\left(\partial^2
    -\mu^2_{{\bf n \, m}}-\frac{8\gamma^2}{r_0^2}\right)V_{i\,\,{\bf n \,
      m}}\right\}, \label{bilinears1}\eea 
where the sum over ${\bf n}$ and ${\bf m}$ is performed over ${\bf
  n}=0,1,2,3,...$ and ${\bf m}=0,\pm 1, \pm 2, \pm 3,...$,
but with the condition $h_{\pm i\,\, 0,0}=0$ and $V_{\pm i\,\,
  0,0}=0$.  Also, as a consequence of the reality conditions
$h_{+i}(X)=h_{-i}^*(X)$, 
$V_{+i}(X)=V_{-i}^*(X)$ and  $V_{i}(X)=V_{i}^*(X)$, we have the relations
$h_{+i\,\,{\bf n \, m}}(x)=h_{-i \,\,{\bf n \, -m}}^*(x)$,
$V_{+i\,\,{\bf n \, m}}(x)=V_{-i \,\,{\bf n \, -m}}^*(x)$ and
$V_{i\,\,{\bf n \, m}}(x)=V_{i \,\,{\bf n \, -m}}^*(x)$. 

In this way, the squared mass operator has finally been transformed
into an algebraic matrix with constant entries and we can find its
eigenvalues exactly.   

\paragraph{6D EYM$\Lambda$ model}

\begin{table}[top]
\begin{center}
\begin{tabular}{|ll|}
\hline
 {\bf 6D Einstein-Yang-Mills-$\Lambda$} \\\hline
 \end{tabular}
\begin{tabular}{|l|l|}
\hline  Squared mass & Multiplicity  \\ \hline &\\
 $0$ & $1$ \\&\\
 $\mu^2_{{\bf n \, m}} +\frac{2 \sqrt2 }{r_0} \mu_{{\bf n \, m}}$ & $1$  \\&\\
$\mu^2_{{\bf n \, m}} -\frac{2 \sqrt2 }{r_0} \mu_{{\bf n \, m}}$ & $1$
 \\ \hline
\end{tabular}\vspace{0.1cm}
\begin{tabular}{|l|}
\hline
 {\bf 6D Supergravity} \\\hline\end{tabular}\begin{tabular}{|l|l|}
\hline  Squared mass & Multiplicity  \\ \hline &\\
$\frac{8}{r_0^2}$ & $1$ \\ &\\
 $\mu^2_{{\bf n \, m}}$ & $2$  \\&\\
$\frac{4}{r_0^2} \left(1+\frac{r_0^2}{4}\mu^2_{{\bf n \, m}} +
  \sqrt{1+r_0^2 \mu^2_{{\bf n \, m}}} \right)$ & $1$\\&\\ 
$\frac{4}{r_0^2} \left(1+\frac{r_0^2}{4}\mu^2_{{\bf n \, m}} -
  \sqrt{1+r_0^2 \mu^2_{{\bf n \, m}}} \right)$ & $1$ 
 \\ \hline
 \end{tabular}
\end{center}\caption{\footnotesize Squared mass KK towers of physical
  spin-1 perturbations around the rugbyball solution to the 6D
  EYM$\Lambda$ model and 6D supergravity, for modes satisfying
  (\ref{validity}). $\mu^2_{{\bf n \, m}}$ is defined in (\ref{mu}),
  but here the KK numbers ${\bf n}, {\bf m}$ run over ${\bf
    n}=0,1,2,3,...$, ${\bf m}=0,\pm 1, \pm 2, \pm 3,...$, with the
  constraint $\{{\bf n},{\bf m}\}\neq \{0,0\}$. 
}\label{table}
\end{table}

\noindent
To address the spin-1 fluctuations in the 6D EYM$\Lambda$ model, we set $\gamma
=0$ and remove the Kalb-Ramond perturbations ($V_{\pm i}=0$) in the above
4D bilinear action, Eq. (\ref{bilinears1}).  By diagonalizing the
corresponding mass-matrix, we find that the explicit helicity-($\pm
1$) towers are as follows
\be M^2_{{\bf n \, m}} = \mu_{{\bf n \, m}}^2\geq 0, \label{helicitypm}\ee
with ${\bf n}=0,1,2,3,...$ and ${\bf m}=0,\pm 1, \pm 2, \pm 3,...$, but
$0 \not< |{\bf m}|\omega \not< 1$ and  
\be M^2_{{\bf n \, m}} = \mu^2_{{\bf n \, m}} \pm \frac{2\sqrt2}{r_0}
\mu_{{\bf n \, m}}\geq 0, \label{spin-1-non-susy}\ee 
with ${\bf n}=0,1,2,3,...$ and ${\bf m}=0,\pm 1, \pm 2, \pm 3,...$, but
$0 \not< |{\bf m}|\omega \not< 1$ and $\{{\bf n},{\bf m}\}\neq
\{0,0\}$. Neither tachyons nor ghosts are found. 
The $\{{\bf n},{\bf m}\}=\{0,0\}$ mode in
(\ref{helicitypm}) is the massless gauge field associated with the 6D
monopole $U(1)$, which we have previously discussed in (\ref{6DU(1)}).
The remaining  modes in (\ref{helicitypm}) are instead the helicity-($\pm$ 1)
components of massive gravitons; we observe that the massive part of
the KK tower (\ref{spin-2-masses}) is exactly reproduced by
(\ref{helicitypm}), according to 4D Poincar\'e invariance. The KK
towers in (\ref{spin-1-non-susy}) correspond instead to physical
spin-1 fields. The complete set of masses for physical spin-1 fields is
given in Table \ref{table}. 

By analyzing those towers, one easily finds that there
are physical massless 4D spin-1 fields (in addition to (\ref{6DU(1)}))
if and only if $\mu^2_{{\bf n \, m}}=8/r_0^2$, which can be restated
as  
\be \{{\bf n},{\bf m}\omega \}= \{1,0\} \quad \mbox{or} \quad \{{\bf
  n},{\bf m}\omega\}= \{0,\pm 1\}. \label{gauge-field}\ee 
Therefore there is at least one 
massless spin-1 field for any value of the tension (this corresponds
to $\{{\bf n},{\bf m}\}= \{1,0\}$). 
In the sphere case ($\omega=1$) we have three ways to satisfy this
condition, that is $$\{{\bf n},{\bf m}\}= \{1,0\}, \{0,\pm 1\},$$ which
correspond to the three
gauge fields of $SU(2)_{KK}$, whereas for positive tension rugbyballs
and generic saddle-spheres
there is only one choice:$$\{{\bf n},{\bf m}\}= \{1,0\},$$
corresponding to the KK
gauge group $U(1)_{KK}$. However, for the special saddle-spheres
for which (\ref{special}) holds, 
{\bf{\it the number of massless vector fields is
enhanced from one plus one to one plus three!}} 

We shall discuss in detail the physical significance of
these modes in the following subsection.

\paragraph{6D supergravity}

\noindent
 We conclude this subsection by providing the helicity-($\pm $1)
 masses for the 6D supergravity (set $\gamma =1$ and keep the
 Kalb-Ramond fluctuations in Eq. (\ref{bilinears1})).  Diagonalizing
 the corresponding mass-matrix, we find:
\be M^2= \frac{8}{r_0^2}> 0 , \label{U(1)monopole}\ee
which is the vector field associated with the monopole $U(1)$,
\be M^2_{{\bf n \, m}} = \mu_{{\bf n \, m}}^2\geq 0, \quad\mbox{with
  multiplicity 3} \label{mass2}\ee 
and
\be M^2_{{\bf n \, m}} = \frac{4}{r_0^2}
\left(1+\frac{r_0^2}{4}\mu^2_{{\bf n \, m}} \pm \sqrt{1+r_0^2
    \mu^2_{{\bf n \, m}}} \right)\geq 0 \label{mass3}\ee 
where ${\bf n}=0,1,2,3,...$, ${\bf m}=0,\pm 1, \pm 2, \pm
3,...$, $\{{\bf n},{\bf m}\}\neq \{0,0\}$ and 
$0 \not< |{\bf m}|\omega \not< 1$. The masses in (\ref{U(1)monopole}), two
towers out of three in (\ref{mass2}) and  
the towers in (\ref{mass3}) correspond to physical spin-1 fields,
whereas one of the towers in (\ref{mass2}) are the helicity-($\pm $1)
components of massive gravitons. The complete set of masses for the
physical spin-1 fields is summarized in Table \ref{table}.  We note
that neither tachyons nor ghosts are found.  

Regarding the massless vector fields the situation is similar to the
6D EYM$\Lambda$ model. It easy to see that the condition for 
masslessness is again (\ref{gauge-field}) and,
therefore, again  we have a single KK massless gauge boson for
positive tensions and generic negative tensions; instead, 
for negative tensions of the form 
(\ref{special}), {\it the number of massless vector fields is enhanced
  from one
  to three!}   

As an effective check of the results presented in this subsection, we have
also derived the aforementioned spectrum in the sphere case ($\omega
=1$) by expanding the bulk fields  over the Wigner functions
as in \cite{sphere} and obtained exactly the sphere limit of our towers.

\subsection{Massless vectors and 4D gauge symmetries}

In the previous subsection we observed {\it three} massless 4D vector
fields amongst the KK spectra for the 6D models\footnote{In the
  EYM$\Lambda$ model there is also a massless vector field
  descending from the higher dimensional $U(1)$ gauge field, which
  forms a 4D $U(1)$ gauge field.} on both the 
sphere and the special saddle-spheres (\ref{special}). 
We shall now address the physical significance of these modes.  One of
them, the one with axial quantum number ${\bf m}=0$, should provide
the gauge boson for the $U(1)_{KK}$ gauge symmetry, descending from the
axial isometry of the internal space.  The other two
massless vectors, having ${\bf m} \neq 0$, are charged under this axial
symmetry and so we may expect the three vectors to fit into a
non-Abelian structure, like $SU(2)$.  For the sphere, this is indeed
the case, and the three massless vectors compose the gauge fields of
an $SU(2)$ gauge symmetry in the 4D theory.  What happens
for the special saddle-spheres, where there is no $SU(2)$ isometry in
the background? Let's take $\alpha = 2,3,\dots$, so we consider the special
saddle-spheres (we also allow for the smooth sphere with $\alpha=1$).

\subsubsection{Why there are three massless vector modes}

Let us begin by understanding why three massless vector modes appear
in the spectrum, despite the fact that any branes clearly break the
$SU(2)$ isometries of the sphere. 

Above, we found that the massless vector
fields arise as a
linear combination of $h^{\,\,m}_{\mu\,\,\,{\bf m}{\bf n}}(x)$ 
and $V_{\mu\, {\bf m}{\bf n}}(x)$ (and $V^{\,\,m}_{\mu\,\,\,{\bf
    m}{\bf n}}(x)$ for 6D supergravity), 
once we have integrated out the extra dimensions.  In detail,
if one takes the squared mass matrix defined implicitly by the
4D bilinear action in (\ref{bilinears1}) in {\it e.g.} the
EYM$\Lambda$ case, one 
finds that the mass eigenstates are ($\{{\bf n},{\bf m}\}\neq (0,0)$):
\bea A_{i\,\,{\bf nm}}&=& \frac{i}{2}h_{+ i\,\,{\bf
    nm}}-\frac{i}{2}h_{-i\,\, {\bf nm}}+
\frac{1}{\sqrt{2}}V_{i\,\,{\bf nm}}, \nonumber\\ U_{i\,\,{\bf nm}}&=&
-\frac{i}{2}h_{+i\,\, {\bf nm}}+\frac{i}{2}h_{- i\,\,{\bf nm}}+
\frac{1}{\sqrt{2}}V_{i\,\,{\bf nm}},\nonumber \\W_{i\,\,{\bf nm}}&=&
\frac{1}{\sqrt{2}}h_{+i\,\, {\bf nm}}+\frac{1}{\sqrt{2}}h_{-i\,\, {\bf
    nm}},\label{masseigenstates}\eea  
corresponding respectively to  $M^2_{{\bf n \, m}} = \mu^2_{{\bf n \,
    m}} - (2\sqrt2/r_0) 
\mu_{{\bf n \, m}}$, $M^2_{{\bf n \, m}} = \mu^2_{{\bf n \, m}} + (2\sqrt2/r_0)
\mu_{{\bf n \, m}}$ and $M^2_{{\bf n \, m}} = \mu_{{\bf n \, m}}^2$ in
(\ref{spin-1-non-susy}) and (\ref{helicitypm}).  Recall that the massless modes
emerge from the $A_{i\,\,{\bf nm}}$ tower, when $\{{\bf n},{\bf m}\omega \}=
\{ 1,0\},\{0,\pm 1\}$.

We can write, then, the harmonic expansion of
$h_{\mu}^{\,\, m}(X)$ as: 
\be
h_{\mu}^{\,\,\,m}(X) = \sum_{{I}=-,0,+} A_{\mu}^{{I}}(x) K^{{I} \,
  m}(\theta,\varphi) + \mbox{massive modes}\, \label{kvfexpansion}.
\ee
Using the expansion (\ref{pm},\ref{pmkk}), the explicit form for the
wave functions (\ref{derivrel}) and the rearrangement in terms of the mass
eigenstates (\ref{masseigenstates}), it is straightforward to show
that $K^{{I} \,m}(\theta,\varphi)$ indeed correspond to the Killing
vectors (\ref{kvf}) on the special saddle-sphere, where ${I}=0$
corresponds to $\{{\bf n},{\bf m}\omega\}=\{1,0\}$ and ${I}=\pm$ to  
$\{{\bf n},{\bf m}\omega\}=\{0,\pm 1\}$.  This is just as in the traditional
KK reduction scheme.

In this way, we confirm that the presence of {\it infinitesimal
  isometries} on the 
internal space, which are generated by
Killing vector fields, is sufficient for the appearance of massless
vector modes -- even if they cannot be integrated to genuine isometries.

\subsubsection{The absence of enhanced gauge symmetries in the full 4D
  theory}

We now ask whether or not these massless vector modes behave as gauge
fields of an $SU(2)$ gauge symmetry.
The linearized 4D
theory cannot probe any non-Abelian structure, and so to understand
the gauge invariance of the full 4D theory, we 
must go beyond linear order.  To this end, we consider a simple
extension of the EYM$\Lambda$ model, where we add a single complex, massless,
neutral
scalar field which has an action:
\be
S_{\Phi} = -\int d^6X \sqrt{-G}\, \partial_M \Phi^* \, \partial^M \Phi \,
\ee
and which assumes a trivial VEV in the saddle-sphere background.
  It
is easy to see that the linearized equation of motion for $Z:=\delta\Phi$
gives rise to the rugbyball scalar harmonics (see above Eq. (\ref{mu})):
\be
Z(X) = \sum_{{\bf m},{\bf n}} z_{{\bf m}{\bf n}}(x) f_{{\bf
    m}{\bf n}}(\theta) e^{i {\bf m} \varphi}
\ee
with the corresponding masses (\ref{mu}):
\be
M^2 = \frac{4}{r_0^2} l \left(l+1 \right) \qquad  \mbox{where}\quad
\,\,l={\bf n} + |{\bf m}|\omega \,.
\ee
The multiplicity of a given mass
is given by $2l+1$ when $l$ is integer or half-odd integer; otherwise
it is given by $2([l]+\frac12) +1$, where $[l]$ denotes the integer
part of $l$.  We also note that for $l$ integer (which corresponds
also to ${\bf m}\omega$ integer), the wavefunction $f_{{\bf
    m}{\bf{n}}}(\theta)$ is an Associated Legendre function, just as
for the spherical harmonics.  The modes with $l$ non-integer are
instead additional harmonics, which generically have no corresponding states
amongst the spherical harmonics nor indeed any of the Wigner functions.     

Now let us ask how the 4D fields $z_{{\bf m}{\bf n}}(x)$ couple to 
the massless vector fields, and in particular if they do in a
gauge-invariant way.  At trilinear level, this coupling descends 
only from the term:  
\be
S(Z^*,h_{\mu m},Z) = -\int d^6X \sqrt{-G} \, \partial^{\mu} Z^*\,
h_{\mu}^{\,\,m} \, 
\partial_m Z \label{trilin6d}
\ee
and its complex conjugate.  
The above trilinear coupling can now be reexpressed in terms of
the 4D fields, and isolating the contributions involving the massless
vectors, Eq. (\ref{kvfexpansion}), we find:
\be
S(z^*, A_{\mu}^{{I}}, z) = -\int d^4x \sqrt{-g_4} \, \partial^{\mu} z^* \,
A_{\mu}^{{I}} \, z' \int d\theta \, d\varphi \, \frac{r_0^2}{4} \,
\alpha \, \sin\theta \, f \, e^{-i{\bf
    m}\varphi}\, K^{{I}\,m} \, \partial_m \left( f'\, 
e^{i{\bf m'}\varphi} \right) \, ,\label{trilin4d}
\ee
where we have suppressed the KK indices $\{{\bf n},{\bf m}\}$ and $\{{\bf
  n'},{\bf m'}\}$ on $z,f$ and $z',f'$ respectively.  Performing the
integral over the internal dimensions:
\be
g_{I}=\int d\theta \, d\varphi  \, \frac{r_0^2}{4} \,
\alpha \, \sin\theta \, f \, e^{-i{\bf
    m}\varphi}\, K^{{I}\,m} \, \partial_m \left( f'\, 
e^{i{\bf m'}\varphi} \right) \, ,\label{overlap}
\ee
we see that the wavefunction overlap (\ref{overlap}) gives the 4D coupling between $z(x)$
and $z'(x)$ {\it via} a massless vector field, $A^{{I}}_{\mu}(x)$:  
\be
 -g_{I}\int d^4x \sqrt{-g_4} \, \partial^{\mu} z^* \,
A_{\mu}^{{I}} \, z' \, .\label{coupling}
\ee

Observe that if the full 4D theory were to respect an $SU(2)$ gauge symmetry
whose gauge fields are the three massless vector modes,
then $z$ and $z'$ would belong to $SU(2)$ multiplets of the same dimension and
(\ref{coupling}) would correspond to the trilinear terms in the gauge invariant combination $-D_{\mu} z^{a*}
D^{\mu} z'^a$, where $D_{\mu} z^a = \partial_{\mu} z^a + A^{{I}}_{\mu} T^{{I}\, 
  a}_{\,\,\,\,\,b} z^b$, the indices $a, b$ run over $a, b=1,\dots, r$ and $r$ is the size of the
multiplet.  For the classic sphere, $\alpha=1$, this is of course the
case, and the wave function overlaps in (\ref{overlap}) are zero
unless $z$ and $z'$ belong to the same $SU(2)$ multiplet, thanks to
the properties of the spherical harmonics.  We shall now see that such
a structure does not hold for the special saddle-spheres.

To this purpose, let us consider the rugbyball harmonics with $0<{\bf
  m}\omega <1$.  The pattern that emerges for the overlaps
(\ref{overlap}) once both
the integrals over 
 $d\theta$ and $d\phi$ are
 performed\footnote{Whilst we have not checked this result for all  
  $\omega$, ${\bf m}$ and ${\bf n}$ the pattern is quite convincing.},
is that a mode, $f$, with $0<{\bf m}\omega <1$ and ${\bf n}$ 
even ({\it respectively} odd) has a non-zero overlap with the
modes, $f'$, for which ${\bf
  m'}\omega = {\bf m}\omega \mp 1$ and all ${\bf n'}$ odd ({\it
  respectively} even).
It can then easily be seen that this prevents the realization of an $SU(2)$
gauge symmetry.  Take for instance the set of modes $\{z\}$ with some
mass-squared $l(l+1)$ in which $0<|{\bf m}|\omega <1$ and 
${\bf n}=0$.   This mass comes only with degeneracy 2, corresponding
to KK numbers $\{0,\pm {\bf m}\}$.  Therefore, if there exists an
$SU(2)$ gauge symmetry, then the modes in $\{z\}$ fall either into an
$SU(2)$ doublet or two singlets.  The overlap (\ref{overlap}) between
the modes $\{0, {\bf m}\}$ and $\{0,-{\bf m}\}$ is zero, and the
subsequent vanishing of the coupling in (\ref{coupling}) tells us that
$\{z\}$ cannot form a doublet.  On the other hand, the modes $\{0,\pm
{\bf m} \}$ do have a non-zero overlap with $\{{\bf n} \,\,
\mbox{odd},\pm {\bf m} - 1/\omega\}$ and $\{{\bf n} \,\,
\mbox{odd},\pm {\bf m} + 1/\omega\}$, and so the two modes in $\{z\}$
each have a trilinear coupling (\ref{coupling}) with towers of $z'$
and the massless vectors fields.  Thus, they cannot be singlets.  In
this way we can conclude that there does not exist an $SU(2)$ gauge
symmetry corresponding to the massless vector fields. 

We would like to draw one more insight into the absence of $SU(2)$
gauge symmetry for the full 4D theory.  The Killing vectors
(\ref{kvf}) can be considered as generators of an $SU(2)$ algebra, and
the mass-squared operator for the saddle-sphere scalars, $-D^2$, can be
understood as the Casimir Operator for the algebra: $-\frac{r_0^2}{4}
\, D^2 = \frac12 \left( K^+ K^- +
K^- K^+ \right) + (K^0)^2$.  The saddle-sphere scalar harmonics form a basis
for the Hilbert space of functions on which the Hermitian operator,
$-D^2$ (plus boundary conditions), acts.  However, the $SU(2)$ ladder
operators, $K^{\pm}$, do not act within this Hilbert space: the action
of $K^{\pm}$ on the harmonics $f_{\bf{mn}}(\theta)e^{i{\bf m}\varphi}$
with $0 < |{\bf m}|\omega < 1$ gives back 
functions which do not obey the NC and HC boundary conditions.  Again, we see
that the saddle-sphere harmonics do not furnish well-defined
representations of the $SU(2)_{KK}$ generated by the Killing vectors, and
it is precisely the modes with $0 < |{\bf m}|\omega < 1$ that are the
problematics ones\footnote{Notice that this range of ${\bf m}$ is empty
for the special saddle-sphere with $\omega = \frac12$ if we impose
the $Z_2$ orbifold projection discussed in Appendix \ref{fullspin0}.  In
this case, 
then, all the KK modes are in well-defined representations of $SU(2)$
(corresponding to the Wigner functions), and we
can expect an $SU(2)$ gauge invariance in the full 4D theory, at least
if we remove the branes and discuss a smooth non-compact 
manifold.  This is not surprising, since --
outside the branes -- the $Z_2$ orbifolding
effectively cancels out the $\delta=-2\pi$ deficit angle, and  we
return to the standard sphere case.}. 

As we have implied above, the absence of an $SU(2)$ KK gauge symmetry in the
4D theory can be understood in the 6D picture as being due
to the absence of a genuine $SU(2)$ isometry in the internal dimensions.

\subsubsection{The emergence of enhanced gauge symmetries at low energies}

Finally, notice that
although the modes with $0 < |{\bf m}|\omega < 1$ do not belong to
well-defined $SU(2)$ representations, the massless wave functions that
we have found are equivalent to those present in the sphere case (up to an
integer constant multiplying $\varphi$) and do furnish well-defined
$SU(2)$ representations\footnote{This is a consequence of the fact
  that in our mass-squared's, $M^2$, as well as in our $f_{{\bf nm}}(\theta)$
  wavefunctions,  ${\bf m}$ and $\omega$ enter only through the
combination ${\bf m}\omega$.  This is obvious for the masses, and
for the wave functions it can be seen from Eq. (\ref{gbetaabcV}),
after setting $\bar\omega=\omega$ to recover the spectra for the
rugbyball.  Furthermore, the massless modes all have integer ${\bf
  m}\omega$. }.  This holds also for the massless spin-2 and spin-1
fields above, as well as the massless spin-0 fields discussed
below\footnote{We should caution that, 
  although there are no symmetries that suggest them to be massless,
  our harmonic analysis has not included the modes with $0 < |{\bf m}|\omega <
  1$ in the spin-1 sector, nor the modes with $0 < |{\bf m}|\omega <
  1$ and $1 < |{\bf m}|\omega <
  2$ in the spin-0 sector.}.  Therefore, the classical low energy 4D
effective theory that 
results from truncating the massive modes does enjoy an $SU(2)$ KK gauge
invariance {\it to all orders in perturbation theory} -- despite the
absence of a genuine $SU(2)$ isometry in the 
extra dimensions.  Indeed, this low energy 4D theory does not
distinguish between a compactification on a smooth sphere or a special
saddle-sphere!

Moreover, we can argue that the above truncation to the massless
sector is a consistent one\footnote{Mathematical consistency may
  of course not be necessary, if the truncation is
  consistent up to some 
  energy scale.}, at least for the bosonic theory that we
have studied whose field content is identical to that of 6D
supergravity.   Then, if we remove the
branes and replace the singular space with a smooth non-compact
manifold,  the local geometry is the same for the 
sphere everywhere and the KK ansatz for the special
saddle-sphere is essentially identical to that of the smooth sphere.
Meanwhile, the sphere reduction of 6D supergravity was shown to be a
consistent one in Ref. \cite{consistent}, thanks to a remarkable
conspiracy between properties of the 2-sphere and the structure of
supergravity.  

\subsection{Massless scalars}\label{0spin-0}

Finally, we turn to the spectra of 4D spin-0 fields, which are
governed by the action given in Appendix \ref{LCG-xi}.  In 
Appendix \ref{fullspin0} we give the complete spectra for unwarped
braneworld compactifications in 6D supergravity.  Here, our focus
shall be on the massless scalars featured in the low energy 4D
effective theory. Again we shall first review the RS model and then
examine the 6D braneworld models. 

\subsubsection{Randall-Sundrum}
In the RS model of Ref. \cite{Randall:1999ee}, the massless scalar
sector involves one normalizable mode (the radion), which becomes non
normalizable in the decompactification limit $r_c\rightarrow \infty$
\cite{radion}. Let us find this mode in our formalism. We can of
course restrict our attention 
to the spin-0 action $S^{(0)}(h,h)$ as in \cite{Randall:1999ee} only
gravity is introduced and the branons are consistently projected out
by the $S^1/Z_2$ orbifold conditions. Therefore, we only have to deal
with the perturbation $h_{\rho \rho}$, because (\ref{h++constraint})
implies $h_{i}^{\,\, i}= -h_{\rho \rho}$. It is easy to derive the EOM
for $h_{\rho \rho}$: 
\be -\frac{1}{\sqrt{-G}}\partial_M\left(\sqrt{-G} \partial^M h_{\rho
    \rho}\right)+ \left[-A'^2 +\frac{1}{3}T \kappa^2 \delta(\rho
  -Y)\right]h_{\rho \rho}=0, \ee 
where $T\kappa^2 \delta(\rho -Y)\equiv T_1 \kappa^2 \delta (\rho) +
T_2 \kappa^2 \delta (\rho - \pi r_c) $. We now perform a KK
decomposition $h_{\rho \rho} (x,\rho) =\sum_{\bf n} h_{\rho \rho}(x)
f_{\bf n}(\rho)$ and focus on the massless case ($\eta^{\mu
  \nu}\partial_{\mu }\partial_{\nu}=0$); we obtain the simple equation 
\be \psi'' = 0,  \label{radion-psi}\ee
where we have defined $\psi \equiv e^A f$ and used property
(\ref{RSproperty}).  The only solution to (\ref{radion-psi})
satisfying the $S^1/Z_2$ orbifold conditions is  $\psi$ constant,
which corresponds to  
\be f\propto e^{-A}. \label{radion-function} \ee
Mode (\ref{radion-function}) is the wave function of the radion. By
inserting this mode in the kinetic term of $h_{\rho \rho}$ in
(\ref{spin-0-hh}) one easily finds that it is normalizable for any
finite $r_c$, but becomes non normalizable in the limit $r_c
\rightarrow \infty$.

\subsubsection{6D Brane Worlds}\label{spin06D}
After this non-trivial check of our formalism we now turn to the
conical-GGP solutions of 6D supergravity.
The stability of the GGP solutions has been investigated in
\cite{Burgess:2006ds} and \cite{Parameswaran:2007cb}, where no
tachyons emerged unless non-Abelian gauge groups are
considered. Indeed, in the presence of non-Abelian gauge groups, an
instability may arise in the sector described by the action
$S^{(0)}(V,V)$, with $V_{\underline{m}}$ orthogonal to the background
monopole\footnote{This instability is also present in the
  sphere-monopole solution \cite{seifinstabilities}, which is a
  particular case of the GGP solutions.}
\cite{Parameswaran:2007cb}. We observe that, even in the absence of
non-Abelian gauge groups, the stability of the GGP solutions is
marginal, in the sense that there are necessarily massless scalars in
the physical spectrum. These massless particles are manifestations of
two symmetries in the model. One is the following invariance of the
EOMs: $G_{MN} \rightarrow w \, G_{MN}$ and $e^{\phi/2} \rightarrow w
\, e^{\phi/2}$, where $w$ is a real number.  Note that this is only a
classical symmetry because the action rescales as $S_B \rightarrow w^2
\, S_B$, so we do not expect the corresponding scalar to remain
massless once quantum corrections are included.  The other is the
Kalb-Ramond symmetry, which acts as $B_2\rightarrow B_2+d\Lambda$, where
$\Lambda$ is a general 1-form field. The actual presence of the zero
mode corresponding to the former symmetry has been shown in
Refs. \cite{Lee:2006ge,Burgess:2006ds}.  

Here, by using our bilinear action, we can easily figure out where the
other massless scalar is. This emerges as the lightest 4D mode of the
field $V_{ij}$, whose bilinear action is simply (see
Eq. (\ref{spin-0-V_2})) 
\be -\frac{1}{16}\int d^6X \sqrt{-G}\, e^{\phi-2A} \partial_M V_{ij}
\partial^M V_{ij}.\ee 
This action is equivalent to the spin-2 action (\ref{spin-2}) in the
case of the conical-GGP solutions, which satisfy $A=\phi/2$. 
The wave functions and mass spectrum coming from $V_{ij}$ are
therefore identical to the one presented in Subsection
\ref{grav-fluct}. For ${\bf n}={\bf m}=0$ we obtain the massless
scalar field associated to the Kalb-Ramond symmetry. 
In the spherical limit this corresponds to the $l=0$ mode in
(\ref{sphere-spin-2}) \cite{Salvio:2007mb}.  

\section{Summary of Results}

Before concluding, let us provide an overview of our results.

\begin{itemize}

\item We have derived the linearized dynamics,
  Eqs. (\ref{spin-2})-(\ref{E:spin-1}) and Appendix \ref{LCG-xi}, for
  the physical  
  perturbations about 
  general backgrounds in a general class of field theories.  In 
  particular, we take Einstein-Yang Mills (EYM) theory in $D$ spacetime
  dimensions, with a bulk dilaton or cosmological constant ($\Lambda$), and a
  number of dynamical 3-branes.  Moreover, for $D=6$ we include a
  dilaton and 2-form
  potential.  Therefore, 6D chiral supergravity, $D$-dimensional
  EYM$\Lambda$ theory and the 5D Randall-Sundrum models all fall
  within our analysis.  The backgrounds considered respect 4D Poincar\'e
  invariance, but may be warped in a radial transverse coordinate.

\item Taking the Randall-Sundrum models as an
  illustrative example within our formalism, we retrieve the
  well-known dynamics for spin-2 
  fluctuations and identify the massless scalar (the radion), which is
  normalizable in the two brane model and becomes non-normalizable in
  the one brane model.

\item For the 6D EYM$\Lambda$ model, we consider the unwarped
  ``rugbyball-monopole'' compactifications, sourced by two
  3-branes of 
  equal tension.  When the
  tensions are zero, we recover the  
  sphere-monopole compactification, and when the tensions are negative we refer
  to the 2D geometry as a ``saddle-sphere''.  By deriving a set of
  ``rugbyball harmonics'', we are able to obtain analytic KK spectra; {\it i.e.} we discuss how to find physical 4D spin-2, spin-1
  and -- consistently truncating branons -- spin-0 fields and their masses.  We 
  present the full spin-2 spectrum and the spin-1 spectra for axial
  momentum number $0\not<|{\bf m}|\omega
  \not<1$.

\item For the 6D supergravity, the backgrounds of interest are the
  warped, axially symmetry braneworld (``conical-GGP'') solutions,
  which have unwarped limits to the rugyballs and saddle-spheres, and
  to the sphere.  Our focus is on the bosonic ``Salam-Sezgin'' sector
  (from the gravity-tensor supermultiplet and the $U(1)$ gauge
  multiplet in which the background monopole lies), since the
  remaining bosonic sectors have been treated elsewhere.  We obtain the 
  complete spin-2 spectrum.  For the spin-1 and spin-0 sectors, we
  restrict to the unwarped backgrounds, and employ the rugbyball
  harmonics to find the spectra.  The sectors covered by our
  analysis\footnote{We also find the spectrum in the sphere case as a
    check.}
  are summarized in detail in Table \ref{T:summary}.

\end{itemize}

\begin{table}[top]
\begin{center}
\begin{tabular}{|l||c|c|c|}
\hline  & spin-2  & spin-1 & spin-0  \\ \hline \hline
 Rugby-ball $\delta \geq 0$ & {\it all modes} & {\it all modes} & {\it
   all modes} \\
 Saddle-sphere $\delta = -2\pi$ & {\it all modes} & {\it all modes} & {\it
   all modes} \\
Generic Saddle-sphere & \phantom{00} {\it all modes} \phantom{00} & ${\bf m}=0$;  $|{\bf m}|\geq
1/\omega$ &  $|{\bf m}|=0,1/\omega$; $|{\bf m}|\geq 2/\omega$
\\
Warped Models & {\it all modes} & -- & -- 
 \\ \hline
\end{tabular}
\end{center}\caption{\footnotesize The sectors covered in the present paper for
  Braneworld Compactifications in 6D Supergravity.  In order to
  address the spin-0 sector, we projected out the branons with an
  orbifolding.  We also here impose the orbifolding for all sectors in the
  presence of negative tension branes. \label{T:summary}}
\end{table}

\noindent
Our main physical results for the 6D braneworlds are as
  follows.  

\begin{itemize}

\item 
The spin-2 spectrum includes the massless 4D graviton
  separated from the rest of the KK tower by a mass gap, and the mass
  gap is indeed observed in all sectors.
  For rugbyballs with positive deficit angles and for generic
  saddle-spheres, the spin-1 sector contains a massless KK gauge boson
  due to the $U(1)$ isometry in the background (in addition to any
  massless 4D gauge bosons descending from unbroken 6D gauge
  symmetries).  
For the special saddle-spheres with deficit angles $\delta = -2\pi,
-4\pi, \dots$, there is a qualitative difference.  Here, there are three
  Killing vectors, which are well-defined everywhere outside the
  branes and obey an $SU(2)$ Lie algebra.  Although only one of them
  integrates to a genuine isometry, the
  number of massless KK vectors fields is consequently enhanced to
  three.  Meanwhile, in the
  spin-0 sector for supergravity, we identify the two massless scalar
  fields expected in all cases from the classical scaling symmetry and the
  Kalb-Ramond symmetry. 

\item The spin-2 and spin-1 spectra are all well-behaved despite the
  presence of codimension-two dynamical brane sources, which induce
  singularities in the bulk geometry. To make
  progress in the spin-0 sector, we had to discard the branon modes
  ({\it e.g.} by placing the branes at orbifold fixed points).

\item  
The spectra analysed -- which incorporates {\em all} modes for rugbyballs
 sourced by positive tension branes -- do not harbour any
 instabilities; neither tachyons nor ghosts.

\item To understand the significance of the three massless 4D vector
  fields that appear for the special saddle-spheres, we go beyond bilinear
  order.  We find that in the full 4D theory, they do {\it
    not} represent gauge fields of an $SU(2)$ gauge symmetry.  This is
  due to the presence of KK modes that are not in well-defined $SU(2)$
  representations.  The classical 
  masslessness of the vector fields is thus not protected by any
  symmetry, which is in accordance with the 
  absence of a genuine $SU(2)$ isometry in the background.  

\item
In the massless sector, however, all modes fall into well-defined
$SU(2)$ representations.  Therefore, the low energy 4D effective theory obtained by truncating to the massless sector does seem to enjoy a classical $SU(2)$ KK gauge symmetry, despite
  the absence of a background $SU(2)$ isometry!  Indeed, this low
 energy effective theory 
 does not distinguish between compactifications on the sphere and the
 special saddle-spheres.    

\end{itemize}

\section{Conclusions}

In this paper, we have provided the dynamics of the physical fluctuations in a
wide class of models, which incorporate the bosonic fields
generically present in bulk supergravity theories -- gravity, non-Abelian
gauge fields, the dilaton and two-form potential -- as well as
dynamical 3-branes.  Our final equations
((\ref{spin-2})-(\ref{E:spin-1}) and those in Appendix \ref{LCG-xi}), 
which can be considered as a generalization of the analysis in
\cite{RandjbarDaemi:2002pq}, provide the starting point to construct a
4D effective field theory emerging from various higher dimensional
models, with compactified extra dimensions and/or branes.

We next proceeded with that objective to study the behaviour of
braneworlds solutions in six dimensions, taking as representative the
rugbyball compactifications of Einstein-Yang Mills theory with a
cosmological constant (EYM$\Lambda$) and 
certain axi-symmetric warped compactifications to 6D minimal gauged
supergravity; the so-called conical-GGP solutions.   We have obtained
the complete KK spectrum for the 4D spin-2 sector in
  the conical-GGP solutions, which is a step towards 
  understanding the behaviour of gravity in codimension two braneworld
  models, as for example
  studied in \cite{gravity}.  The spin-1 and spin-0 sectors
present large systems of coupled differential equations to be solved
(five-by-five 
for the  spin-1 fluctuations, eight-by-eight for the spin-0 fluctuations
after truncating the branons), and we are
able to do so in the unwarped cases by developing ``rugbyball
harmonics'', in analogy to the spherical harmonics.  Along the way, we
also recovered some familiar features of the 5D Randall-Sundrum
models.  Our main results are summarized in the previous section.

Previous studies have revealed that codimension-two braneworld
compactifications can evade the traditional KK lore in several
ways.  For instance, in \cite{Parameswaran:2006db} it was found that
the KK mass-gap can be decoupled from the size of the extra
dimensions in the presence of negative tension branes, in principle
allowing not only gravity but also the SM to propagate in large extra 
dimensions.  This phenomenon can also be observed here.  We can also
now suggest the following.  The power-law warping present in the 6D
braneworlds studied here does not change qualitatively the physics.
Moreover, models with only positive tension codimension-two branes also have
qualitatively the same behaviour as traditional KK
compactifications.  Meanwhile, the introduction of negative tension
codimension-two branes can lead to surprising dynamics.

As yet another
example of how the
physics of braneworlds in 6D can counter intuition, we have found -- for
special saddle-sphere compactifications with deficit angles $\delta =
-2\pi, -4\pi, \dots$ -- three massless vector
fields thanks to the presence of three $SU(2)$ Killing vectors in
the internal manifold that are well defined everywhere outside the
branes.  Thus we see that infinitesimal isometries are sufficient to
imply massless vector fields, even if they cannot be integrated to
genuine isometries.  All the massless modes in the models studied here
fall into well-defined 
representations of the $SU(2)$, although there are massive KK modes which
do not.  In this way we see that the massless vectors provide the
gauge fields of an 
enhanced $SU(2)$ KK gauge symmetry in the classical, low
energy, 4D effective theory obtained by truncating to the massless
sector, despite the absence of an $SU(2)$ 
isometry in the background!  Apparently, the low energy theory does
not distinguish between a compactification on the special saddle-spheres and
the smooth sphere.  

At the same time, as we approach the energy of the
KK mass gap and incorporate the non-zero modes, we see
that the $SU(2)$ KK gauge symmetry is broken explicitly to $U(1)$.
This is because only the $U(1)$ is a genuine global continuous
isometry of the internal manifold.  The
masslessness of the extra massless vector fields is thus not protected by
any symmetry, and should not survive quantum corrections.  Meanwhile, reaching
energies far above the KK mass-gap, the full 6D symmetries will be
restored as usual.  The pattern of symmetry breaking and emergence
that we have found within our classical approximation, as different
energy scales are probed, is thus a novel one.   

In the model whose field content and structure
corresponds to the bosonic part of 6D supergravity, the low-energy
theory describes the graviton, the three vectors in the adjoint of
$SU(2)$ and two massless scalars that are $SU(2)$ singlets.    Whether
the above properties are shared with fermionic modes is not known
and their behaviour, though of interest, lies beyond the scope of the
present paper.  Meanwhile, we argued that we expect the zero-mode truncation 
to be a consistent one, at least in the aforementioned model once we
remove the brane sources and study a non-compact smooth manifold. We
thus note that this bosonic model is in principle a complete one,
sufficient to demonstrate the unconventional dynamics that we have
observed. It would certainly be interesting to check the 
consistency also in the presence of branes.  

This work concludes our study of the bosonic perturbations in the
axi-symmetric brane-world solutions to 6D supergravity.  We may now
turn to the fermionic sector.

\vspace{0.7cm}

{\bf Acknowledgments.}   It is a pleasure to thank Cliff Burgess,
Claudia de Rham, Stephan Mohrdieck, 
Nicola Pagani, Riccardo Rattazzi, Michele Redi, Mikhail Shaposhnikov and Andrea
Wulzer for valuable discussions.   S.L.P. is
supported by the Deutsche Forschungsgemeinschaft under the
Collaborative Research Center 676 and by the 
European Union 6th framework program MRTN-CT-503359 ``Quest for
Unification''.  A.S. has been 
supported by the Tomalla Foundation and by CICYT-FEDER-FPA2008-01430.
S.L.P. and A.S. thank the High 
Energy Theory Group at ICTP for hospitality at various stages of this work. 

\vspace{0.7cm}

\appendix

{\Large \bf Appendix}

\section{General $\xi$-Dependent Bilinear Action}\label{general-xi}

Here we give the explicit expression for the biliner action that
depends on the fluctuations of the brane positions $\xi^M$, before any
gauge fixing, that is the 
last two terms in (\ref{split}). These terms have been computed by
varying the brane action (\ref{Sb}) with respect to
(\ref{perturbations}) and by keeping only terms up to the quadratic
order. Their explicit expression is the following 
\bea S(\xi, \xi )&=& -\frac{T}{2} \int d^4x \sqrt{-g} \left[ G_{MN}
  \partial \xi^M \cdot \partial \xi^N \right. \nonumber \\ &&
 +\frac{1}{2} \xi^P \xi^R \partial_P \partial_R G_{MN} \partial Y^M
 \cdot \partial Y^N +2\xi^P \partial_P G_{MN}  \partial \xi^M\cdot
 \partial Y^N \nonumber \\ && + \frac{1}{2} \xi^P
 \partial_P G_{MN} \, \xi^R \partial_R G_{SQ} \, \left( \frac{1}{2}
   \partial Y^M \cdot \partial Y^N \partial Y^S \cdot \partial Y^Q -
   \partial Y^M \cdot \partial Y^S \partial Y^N \cdot \partial
   Y^Q\right) \nonumber \\   
&& + G_{MN} G_{PR}  \left(  \partial \xi^M \cdot \partial Y^N \partial
  \xi^P \cdot  \partial Y^R -2\partial \xi^M \cdot \partial \xi^P
  \partial Y^N \cdot \partial Y^R\right)  
\nonumber \\ &&\left. + \xi^P \partial_P G_{MN} G_{RS} \left( \partial
    Y^M \cdot \partial Y^N \partial \xi^R \cdot \partial Y^S  
-2 \partial Y^M \cdot \partial \xi^R \partial Y^N \cdot \partial Y^S \right)
\right], \label{xi-xi}\eea
and 
\bea 
S(h,\xi)&=& -\frac{T}{2} \int d^4 x \sqrt{-g}\left[\xi^P \partial_P
  h_{MN} \partial Y^M \cdot \partial Y^N +2 h_{MN} \partial \xi^M
  \cdot \partial Y^N \right. \nonumber \\ && + h_{MN} \xi^P\partial_P
G_{RS} \left(\frac{1}{2} \partial Y^M \cdot \partial Y^N \partial Y^R
  \cdot\partial Y^S -\partial Y^M \cdot \partial Y^R \partial Y^N
  \cdot \partial Y^S\right)  \nonumber \\ && 
\left.  + h_{MN} G_{PR} \left( \partial Y^M \cdot \partial Y^N \partial
    \xi^P \cdot \partial Y^R -2 \partial Y^M \cdot \partial  \xi^P
    \partial Y^N \cdot \partial Y^R \right)\right]. \label{h-xi}\eea 
The bulk quantities in (\ref{xi-xi}) and (\ref{h-xi}), that is the
background metric $G_{MN}$ and the fluctuation $h_{MN}$, are computed
in the background brane position. This is because (\ref{xi-xi}) and
(\ref{h-xi}) come from the variation of the brane action (\ref{Sb})
where the bulk fields are computed in the brane position.

\section{Spin-0 Bilinear Action in the Light Cone Static Gauge}\label{LCG-xi}

Here we provide the spin-0 action in the light cone static gauge
defined by (\ref{LCG2}) and (\ref{static}). This is the only part
where the branons $\xi^{\underline{m}}$ appear. 

Let us start with the spin-0 action that only depends on the bulk
fields. The non vanishing terms are the following:
\bea S^{(0)}(h,h) &=& -\frac{1}{4\kappa^2}\int d^DX \sqrt{-G} \left[
  \partial_{\mu}h_{\underline{mn}} \partial^{\mu}h^{\underline{mn}} +
  \partial_{\rho}h_{\underline{mn}}\partial_{\rho}h^{\underline{mn}}
  +h_{\underline{mn};l}h^{\underline{mn};l}  \right. \nonumber \\&& +
h_{\rho \rho}^2(D_2 A' B' +2A'')+2 \left(A'' + A'^2\right) h_{\rho
  \rho} h_i^{\,\,\, i}\nonumber \\ 
&&\left(D_2A'B'+ 2A''-\frac{1}{2}B'^2-2B'' \right)h_{\rho
  m}h_{\rho}^{\,\,\, m}-4A' h_{\rho}^{\,\,\,
  \underline{n}}h_{m\underline{n}}^{\quad \, ;m} \nonumber \\
&&+h_i^{\,\,\, i}h_m^{\,\,\, m}A' B' +2 \left( B''+
  \frac{B'^2}{2}\right)h_{\rho \rho} h_{m}^{\,\,\,\, m}+\frac{1}{2}
B'^2 h_{mn} h^{mn} +\frac{1}{2}B'^2 \left(h_m^{\,\,\,
    m}\right)^2\nonumber \\ &&-2 e^{-B}
h^{m}_{\,\,\,\,l}h^{n}_{\,\,\,\, h} \Omega_{mn}^{\quad
  \,\,lh}+2\kappa^2 h_{\underline{lm}}h^{\underline{l}}_{\,\,\,
  \underline{n}}\left( \frac{1}{2} e^{\phi/2} F^{\underline{m}}_{\quad
    \underline{h}}F^{\underline{nh}} +\frac{1}{4\kappa^2}
  \partial^{\underline{m}} \,\phi \,\partial^{\underline{n}} \,
  \phi\right)\nonumber \\ 
&&+\kappa^2 e^{\phi/2}
h^{\underline{mn}}h^{\underline{lh}}F_{\underline{lm}}F_{\underline{hn}}
+ \frac{1}{2}
\left(\partial_{\mu} h_{i}^{\,\,\, i}\partial^{\mu} h_j^{\,\,\, j} 
  +\partial_{\rho}h_{i}^{\,\,\, i}\partial_{\rho} h_j^{\,\,\, j}+
  h_{i\,\,\, ; m}^{\,\,\, i}h_j^{\,\,\, j;m}\right) 
\nonumber \\  
&&\left. + \left(h_i^{\,\,\,i} \right)^2 \left(\frac{1}{2} A'^2
    +\frac{T}{2}\kappa^2 \sqrt{g/G}\,\,\delta(X_c- Y_c)\right)
\right], \label{spin-0-hh}\eea 
where $\Omega_{mn\,\,\, h}^{\quad \,\,l}$ is the Riemann tensor for
the metric $K_{mn}$ and $X_c$ and $Y_c$ are defined below
Eq. (\ref{spin-1-hh}). We observe that in the last line of
(\ref{spin-0-hh}) there is an explicit brane contribution (the tension
of the brane $T$ appears explicitly). Moreover, 
\bea S^{(0)}(V,V) &=& -\frac{1}{2}\int d^DX
\sqrt{-G}\,e^{\phi/2}\left[ \partial_{\mu} V_{\underline{m}}
  \partial^{\mu} V^{\underline{m}} 
+ D_{\underline{m}} V_{\underline{n}} D^{\underline{m}}
V^{\underline{n}} + \left(-2 A'^2 +\frac{1}{4}
  \phi'^2\right)V_{\rho}^2  \right. \nonumber \\  
&&+\left(-2A' + \phi'\right)V_{\rho}
D_{\underline{m}}V^{\underline{m}}+R^{\underline{mn}}V_{\underline{m}}
V_{\underline{n}}
+2\,\overline{g}F^{\underline{mn}}V_{\underline{m}}\times
V_{\underline{n}}\nonumber \\&& \left.-\frac{1}{2}\phi' V_{\rho} D_m
  V^m+\frac{1}{2}\phi' V^{m} D_m V_{\rho} +\kappa^2 e^{\phi/2}
  \left(F_{\underline{l}}^{\,\,\,
      \underline{m}}V_{\underline{m}}\right)
  F^{\underline{ln}}V_{\underline{n}}\right],  \\ 
S^{(0)}(h,V) &=& \int d^DX \sqrt{-G}
e^{\phi/2}\left[F^{\underline{nm}}V_{\underline{m}}\,h_{\underline{ln}}^{\quad
    ;\underline{l}} 
+\left(D_{\underline{n}}V_{\underline{m}}-
  D_{\underline{m}}V_{\underline{n}}\right)h_{\underline{l}}^{\,\,\,\underline{n}} F^{\underline{lm}}\right.\nonumber \\ &&\left. -\frac{1}{2} A' F^{\rho \underline{m}} V_{\underline{m}} h_i^{\,\, i} + A' F^{\underline{mn}} V_{\underline{n}} h_{\underline{m} \rho}\right],\\    
S^{(0)}(\tau, \tau)&=& \hspace{-0.2cm}-\int d^DX  \sqrt{-G}
\left[\frac{1}{4\kappa^2} \partial_M\tau \,\partial^M\tau +\frac{1}{2}
  \left(\frac{\partial^2 \mathcal{V}}{\partial
      \phi^2}+\frac{1}{16}e^{\phi/2}F^2+\frac{\phi'^2}{4\kappa^2}\right)\tau^2\right], \\  
S^{(0)}(h,\tau)&=& \int d^DX
\sqrt{-G}\left\{\frac{1}{2\kappa^2}\left[\phi' \tau \left(A' h_{\rho
        \rho} + h_{\underline{m} \rho}^{\quad ;
        \underline{m}}-\frac{1}{2}A'
      h_i^{\,\,i}\right)+h^{\underline{m}\rho}
    \partial_{\underline{m}} \tau \,\phi' \right]\right. \nonumber \\ 
&&\left.  +\frac{1}{4}e^{\phi/2} F^{\underline{ml}}
  \,F^{\underline{n}}_{\,\,\,\,\, \underline{l}}\, \tau
  \,h_{\underline{mn}}\right\},\\ 
S^{(0)} (V, \tau) &=& \int d^DX \sqrt{-G}
\,e^{\phi/2}\left[\frac{1}{4} F^{\underline{mn}} \,\tau
  \left(D_{\underline{n}} V_{\underline{m}} 
-D_{\underline{m}} V_{\underline{n}}\right)-\frac{1}{2} \phi' F^{\rho
m} \tau V_m\right],\\ 
S^{(0)}(V_2, V_2) &=& -\frac{1}{16}\int d^DX \sqrt{-G}\,
e^{\phi}\left\{e^{-2A} \partial_M V_{ij} \partial^M
  V_{ij}\right.\nonumber \\ 
&&-2e^{-4A -2\phi} 
\left(e^{\phi +2A} V^{\underline{n}}_{\quad
    \underline{m}}\right)_{;\underline{n}} \left(e^{\phi +2A}
  V^{\underline{lm}}\right)_{;\underline{l}} 
\nonumber \\ &&-4V^{\underline{mn}}
e^{-A}\partial_{\underline{m}}\left[e^{-A-\phi} \left( e^{\phi +2A}
    V^{\underline{l}}_{\,\,\,\underline{n}}\right)_{;\underline{l}}\right] \nonumber \\ &&\left.+\partial_{\mu} V_{\underline{mn}}\partial^{\mu}V^{\underline{mn}}+\frac{1}{3} V_{[\underline{nl};\underline{m}]}V^{[\underline{nl};\underline{m}]} +\frac{\kappa^2}{2}\gamma^2 e^{\phi/2} \left(V_{\underline{mn}} F^{\underline{mn}}\right)^2\right\}, \label{spin-0-V_2}\\   
S^{(0)}(V,V_2)&=&-\frac{\kappa}{4} \gamma\int d^DX \sqrt{-G}\,
e^{\phi} V_{\underline{mn}} F^{\underline{mn}} \left[
  \left(A'+\frac{1}{2}\phi'\right)V_{\rho}
  +D_{\underline{l}}V^{\underline{l}}\right]. 
\eea
We have no mixing of the form $S^{(0)}(h, V_2)$ and $S^{(0)}(\tau,
V_2)$ as a consequence of $H_{MNP}=0$ (at the background level). We
have checked that the term $S^{(0)}(V,V)$ reduces, as it should, to
the corresponding action in Ref. \cite{Parameswaran:2007cb} in the
case in which $V_{\underline{m}}$ is orthogonal to the background
gauge field.   

 Let us consider now the branon-dependent action. This turns out to
 have the following form\footnote{The term of the form $S^{(0)}(V_2,
   \xi)$ vanishes as a consequence of $H_{MNP}=0$ at the background
   level, which in turn follows from our background ansatz.}: 
\be S^{(0)}(h, \xi) + S^{(0)}(V, \xi) +S^{(0)}(\tau , \xi)+S^{(0)}(\xi,\xi).\ee
 Therefore, the fields $\xi^{\underline{m}}\,$ in general couple with
 some bulk fields, but these mixings are confined  
to the spin-0 action. The explicit expressions for the different pieces are
\bea S^{(0)}(h, \xi)&=& -\frac{T}{2} \int d^4x \sqrt{-g}
\left[\,2\,\xi^{\underline{m}} \left(A' h_{\rho \underline{m}}+
    h_{\underline{nm}}^{\quad \, ;\underline{n}}\right) +e^{-A}
  \xi^{\underline{m}}\partial_{\underline{m}} h_{ii}\right], \\ 
S^{(0)}(V, \xi)&=& T\kappa^2 \int d^4x\, \sqrt{-g}\, e^{\phi/2}
F_{\underline{m}}^{\,\,\, \,\,\underline{n}}\, V_{\underline{n}}\,
\xi^{\underline{m}},\\  
S^{(0)}(\tau , \xi)&=& \frac{T}{2}\int d^4 x \sqrt{-g}\,\,
\xi^{\underline{m}} \,\partial_{\underline{m}} \phi \, \,\tau, \\ 
S^{(0)}(\xi,\xi)&=& -\frac{T}{2} \int d^4x \sqrt{-g}
\left[G_{\underline{mn}} \partial_{\mu} \xi^{\underline{m}}\,
  \partial^{\mu} \xi^{\underline{n}}+\frac{1}{2} \xi^{\underline{m}}
  \partial_{\underline{m}} g_{\mu \nu} \, \xi^{\underline{n}}
  \partial_{\underline{n}} g_{\eta \sigma} \, \left(\frac{1}{2} g^{\mu
      \nu} g^{\eta \sigma} - g^{\mu \eta}g^{\nu
      \sigma}\right)\right.\nonumber \\ 
&&+\frac{1}{2} \xi^{\underline{m}}\xi^{\underline{n}}
\,\partial_{\underline{m}} \partial_{\underline{n}} g_{\mu \nu } \,
g^{\mu \nu}  \left. 
+ T\kappa^2\sqrt{g/G} \, \delta(Y_c - Y_c) \xi^{\underline{m}}
\xi_{\underline{m}}\right].\label{spin-0-xi-xi}\eea 

We discuss the various singularities that can be observed in the
above in Subsection \ref{spin-0-action} and below.

\section{Spin-0 Spectrum  for 6D Supergravity
  Compactification}\label{fullspin0} 

We finally analyse the (massive) spin-0 fluctuations in 6D braneworlds
by using the general spin-0 action given in Appendix 
\ref{LCG-xi}. Here we discard the branons. There are different ways
to make this truncation consistently, {\it e.g.} by introducing an
orbifold that 
projects them out. In Ref. \cite{Parameswaran:2007cb} such an orbifold
has been defined taking into account the presence of at least two
patches in the description of spherical topologies.  Here we only use
the fact that the orbifold action in the intersection of the two
patches is $\varphi \rightarrow \varphi + \pi$.  
 In the absence of the branons the $\delta(0)$ singularities mentioned
 in Subsection \ref{spin-0-action} obviously disappear. We shall see
 that it is also possible to deal with the other type of singularities
 mentioned there and extract a finite spectrum. 

Here we focus on the unwarped solutions and in particular on the
rugbyballs and saddle-spheres
defined in Subsection \ref{EOM}. In this case we will be
able to generalize the harmonic analysis developed in Subsection 
\ref{6D-brane-1} to the spin-0 sector, which involves 2D tensors as
well as 2D vectors and scalars.  This technique allows us to
transform complicated coupled differential equations into algebraic
equations whose solutions can be found exactly. The relevant
fluctuations are   
$h_{mn}$, $V_{m}$, $\tau$, $V_{ij}$
and $V_{mn}$, where $m$ and $n$ run over $\theta$ and $\varphi$. We
observe that the fluctuations $V_m$ orthogonal to background gauge
field decouple to the other fields and have already been analyzed in
\cite{Parameswaran:2007cb}; therefore here we assume $V_m$ to be
parallel to the background gauge field. One should keep in mind that,
if the branons are projected out by the above-mentioned orbifold,
only the modes with ${\bf m}$ even survive (in the Fourier expansion
over $e^{i {\bf m}\varphi}$). The spin-0 action in the light cone
gauge assumes the following form: 
\bea S^{(0)}(h,h) &=& -\frac{1}{4\kappa^2} \int d^6X \sqrt{-G}
\left[\partial_{\mu}h_{mn} \partial^{\mu} h^{mn}+
  h_{mn;l}h^{mn;l}\right.\nonumber \\ && -2
h^{m}_{\,\,\,\,l}h^{n}_{\,\,\,\,h}R_{mn}^{\quad \, lh}+ \kappa^2
h_{lm}h^l_{\,\,\,\,n}F^m_{\,\,\,\,\,h}F^{nh}+ \kappa^2
h^{mn}h^{lh}F_{lm}F_{hn}\nonumber \\
&&\left. -\frac{1}{2}h_m^{\,\,\,\,m}(\partial^2 + D^2) h_n^{\,\,\,\,n}
  +
  \frac{T}{2}\kappa^2(h_m^{\,\,\,\,m})^2\sqrt{g/G}\,\delta(X_2-Y_2)\right],\nonumber \\   
S^{(0)}(V,V)&=&-\frac{1}{2}\int d^6X \sqrt{-G} \left[\partial_{\mu}V_m
  \partial^{\mu}V^m + V_{m;n}V^{m;n}\right. \nonumber \\
&&\left. +\frac{1}{2} R V_mV^m + \frac{\kappa^2}{2}F^2 V_m V^m\right],
\nonumber \\ S^{(0)}(\tau,\tau)&=&\frac{1}{4\kappa^2}\int d^6X
\sqrt{-G}\left[\tau \left(\partial^2 +D^2 -\frac{4}{r_0^2}\right)\tau
\right], \nonumber \\ 
S^{(0)}(V_2,V_2) &=& \int d^6X \sqrt{-G}\left[ \frac{1}{16}V_{ij}
  (\partial^2 + D^2) V_{ij}+\frac{1}{16} V_{mn}(\partial^2 + D^2)
  V^{mn}\right. \nonumber \\ && \left. -\frac{\kappa^2 \gamma^2}{32}
  (V_{mn} F^{mn})^2\right],\nonumber \\ S^{(0)}(h,V)&=& \int d^6X
\sqrt{-G}\left[-V_{n;m} h_l^{\,\,\,\, n} F^{lm} \right],\nonumber \\
S^{(0)}(h,\tau) &=& \frac{1}{8} \int d^6X \sqrt{-G}\,F^2 \,\tau\,
h_m^{\,\,\,\,m}, \nonumber \\ S^{(0)}(V,\tau)&=&  \int d^6X
\sqrt{-G}\,\frac{\tau}{4}F^{mn}(V_{m;n} -V_{n;m}), \nonumber \\
S^{(0)}(V,V_2) &=&-\frac{\kappa}{4}\gamma\int d^6X
\sqrt{-G}\,V_{mn}F^{mn}V_l^{\,\,\,\,;l}, \label{spin-0-lc}\eea 
where we have used the light cone gauge relation $h_i^{\,\,\,\,i}+
h_m^{\,\,\,\,m}=0$ in $S^{(0)}(h,h)$ and the property
$F^{ml}F^{n}_{\,\,\,\, l} = G^{mn} F^2/2$ in $S^{(0)}(V,V)$ and
$S^{(0)}(h,\tau)$, which  is a consequence of (\ref{spin-1-system}). 
 
We now want to use a technique similar to that explained in the spin-1
sector, in order to transform the above differential problem into an
algebraic one. Note that the
method provided in Subsection 
\ref{6D-brane-1} can be already applied to perform this transformation
in the terms $S^{(0)}(V,V)$, $S^{(0)}(\tau, \tau)$,
$S^{(0)}(V_2,V_2)$, $S^{(0)}(V,\tau)$ and $S^{(0)}(V,V_2)$ as they
only involve 2D scalars and 2D vectors\footnote{The analysis of the
  $V_{mn}$-EOMs shows that $V_{mn}/\sqrt{-G}$ is a 2D scalar.}. 
What we have done there is to identify
appropriate mass-squared operators from the diagonal part of the
bilinear action, which are Hermitian once the HCs
are imposed.  In this way we were able to define complete sets of 
2D scalar and vector harmonics.  Then we focused on the cases in which
the derivative 
 relation, Eq. (\ref{derivrel}), between scalar and vector harmonics
 holds.  That relation is what allowed
us to deal with the derivative couplings between scalars and vectors
and transform the spin-1 differential problem into an algebraic one,
which could easily be solved.

Here we generalize the above procedure to include the 2D tensor
fluctuations in $h_{mn}$. 
Indeed, $h_{mn}$ can be decomposed into its trace, $h_m^{\,\,\,\,m}$, and
traceless part, $\tilde{h}_{mn}\equiv h_{mn} -
G_{mn}h_l^{\,\,\,\,l}/2$, so that the first entry in
(\ref{spin-0-lc}) decomposes into the two terms:
\bea S^{(0)}(h_m^{\,\,\,\,m},h_m^{\,\,\,\,m}) &=& \frac{1}{8\kappa^2} \int
d^6X \sqrt{-G} 
\left[h_m^{\,\,\,\,m}\left(\partial^2 + D^2 - \frac{\kappa^2}{4}
    F^2\right)h_n^{\,\,\,\,n}\right], \nonumber\\
S^{(0)}(\tilde h_{mn},\tilde h_{mn}) &=& \frac{1}{4\kappa^2} \int
d^6X \sqrt{-G} 
\left[\tilde h_{mn}\left(\partial^2 + D^2 - R\right)\tilde h^{mn}\right],
\eea
where we used the following identities:
\be R_{pmqn}=\frac{R}{2}(G_{pq}G_{mn}-G_{mq}G_{pn}), \quad F_{lm}
F_{hn}=\frac{F^2}{2} (G_{lh} G_{mn}-G_{mh}G_{ln}). \ee 
Observe that $h_m^{\,\,\,\, m}$ is in fact
a 2D scalar field, and we can expand it in terms of the 2D scalar
harmonics found in Subsection \ref{6DBraneWorlds}.  The fluctuations
$\tilde{h}_{mn}$ are instead genuine 2D tensor fluctuations, and the
appropriate mass-squared operator is $-D^2+R$.  Thus, we would like to
solve the eigenproblem:
\be \left(-D^2+R\right)\tilde{h}_{mn}=\mu_{\T}^2 \, \tilde{h}_{mn},
\label{htildeEOMs} \ee 
with the given NCs and HCs, where $\mu_{\T}^2$ are the corresponding
mass-eigenvalues.

In a general basis, Eq. (\ref{htildeEOMs}) is a set of two
coupled differential equations (the traceless property removes one out
of the three components of a rank two symmetric tensor in two
dimensions). However, by writing down Eq. (\ref{htildeEOMs}) in the
$\pm$ basis defined in (\ref{pm}):  
\be \left(-D^2+ R \right)h_{\pm \pm}=\mu_{\T}^2 \, h_{\pm \pm} ,
\label{htildeEOMs2}\ee 
where we used $h_{\pm \pm}=\tilde{h}_{\pm \pm}$, and by explicitly
evaluating $ D^2h_{\pm \pm}$, one finds that the equations for $h_{++}$ and
$h_{--}$ are decoupled, like those of the $h_{+i}$ and $h_{-i}$ fields
in the spin-1 sector. After a long but straightforward calculation we
find 
\be -\partial_{\theta}^2 f^{\pm \pm} + \frac{\dot{B}}{2} \partial_{\theta}
f^{\pm \pm} + \left({\bf m}^2e^{-B}\pm 2{\bf m}\dot{B}
  e^{-B/2}+\frac{\dot{B}^2}{2}-\frac{\ddot{B}}{2}\right) f^{\pm \pm} =
\frac{r_0^2}{4}\, \mu_{\T}^2
f^{\pm \pm},\label{++EOM}\ee 
where a dot represents a derivative with respect to $\theta$ and
$f^{\pm \pm}$ is the wave function of $h_{\pm \pm}$, 
defined by a KK expansion 
\be h_{\pm \pm}(X)=\sum_{{\bf n},{\bf m}}h_{\pm \pm \,{\bf nm}}(x) f^{\pm
  \pm}_{{\bf nm}}(\theta)e^{i {\bf m}\varphi}, \quad \mbox{with} \,\,
{\bf m}=\mbox{generic integer} \ee  
and in (\ref{++EOM}) the KK numbers ${\bf n}$ and ${\bf m}$ are
understood.  The eigenvalues $\mu_{\T}^2$ can be found by using the
technique discussed in Ref. \cite{Parameswaran:2006db}: one can put
the equations into the hypergeometric form, consider the general solution to
the hypergeometric equation and then impose the HCs and NCs. We find 
\begin{itemize}
\item For $|{\bf m} | \omega \geq 2$
\be \mu_{\T}^2=\frac{4}{r_0^2}\left[({\bf n}+|{\bf m} | \omega)({\bf
    n} +|{\bf m} | \omega+1)-2\right]\label{diagonal1}\ee 
\item For $-2< {\bf m}\omega< 2$
\be \mu_{\T}^2=\frac{4}{r_0^2}\left[({\bf n}+2)({\bf n}
  +3)-2\right]\label{diagonal2}\ee 
\end{itemize}
where ${\bf n}=0,1,2,3,...\,$.   In this way we have found a complete
set of 2D tensor harmonics.

We now remember that, in the spin-1 sector analysed in Subsection
\ref{6D-brane-1}, one can generate the 2D vector harmonics 
by acting with derivatives over the 2D scalar harmonics 
(see Eq. (\ref{derivrel}) and the discussion right
above). We can imagine that something similar happens here and
the 2D tensor harmonics (\ref{htildeEOMs2}) can be
generated by acting with derivatives over 2D vector harmonics.
This is indeed the case and in order to see it let us
consider the 2D vector harmonics for $V_m$: 
\be \left(-D^2+\frac{R}{2}\right)V_m =\mu_{\V}^2 V_m,\label{VmEOM}\ee 
where $\mu_{\V}^2$ are the 2D vector mass-eigenvalues.  From now on
we shall assume Condition (\ref{validity}), so that $\mu_{\V}^2=\mu^2$,
with $\mu^2$ the 2D scalar mass-eigenvalues given in (\ref{mu}).
After some manipulation it is easy to show that if
$V_m$ satisfies the previous equation then we also have 
\be -D^2\tilde{V}_{m;n} + R(\tilde{V}_{m;n}+\tilde{V}_{n;m})+
\frac{1}{2} (R_{;m}V_n + R_{;n} V_m - G_{mn} R_{;l}V^l)=\mu^2
\tilde{V}_{m;n}, \label{V;mn1}\ee  
where $\tilde{V}_{m;n}\equiv V_{m;n} -G_{mn}V_l^{\,\,;l}/2$. This
equation is valid for any unwarped compactification, but in the
rugbyball case it can be simplified. Although the Ricci scalar is not
constant everywhere like in the sphere limit as it contains
delta-functions, these additional delta function terms can be
discarded in Eq. (\ref{V;mn1}) because they are dominated by stronger
singularities\footnote{This is a quite generic property of rugbyball
  compactifications \cite{Parameswaran:2006db,Parameswaran:2007cb}.},
which emerge from $D^2\tilde{V}_{m;n}$. This allows us to write
(\ref{V;mn1}) as follows: 
\be -D^2\tilde{V}_{m;n} + R_s(\tilde{V}_{m;n}+\tilde{V}_{n;m})= \mu^2
\tilde{V}_{m;n}, \ee 
where $R_s$ is the Ricci scalar of the sphere ($R_s=8/r_0^2$), or, in
the $\pm$ basis, 
\be \left(-D^2+ 2R_s \right) V_{\pm;\pm}= \mu^2 V_{\pm;\pm},\label{V;mn2}\ee
where we used $V_{\pm;\pm}=\tilde{V}_{\pm;\pm}$. 
Now, comparing the eigenproblems for $h_{\pm\pm}$ and $V_{\pm;\pm}$,
Eqs. (\ref{htildeEOMs2}) and (\ref{V;mn2}), we see that their
eigenfunctions will belong to the same orthogonal set provided that:
\be
\mu_{\T}^2 = \mu^2 - R_s = \mu^2 - 8/r_0^2 \,. \label{muT/V}
\ee
By comparing the 2D vector mass-eigenvalues, $\mu^2$ given in
(\ref{mu}),  with the 2D tensor eigenvalues, $\mu_{\T}^2$ given in
Eqs. (\ref{diagonal1}) and (\ref{diagonal2}), we find that Condition
(\ref{muT/V}) is indeed true in the following cases: 
\begin{itemize}
 \item For ${\bf m }=0$ or $|{\bf m }|\geq 2/\omega$, which we denote
   by $0\not<|{\bf m}|\omega 
  \not<2$, with the constraint $\left\{{\bf
    n},{\bf m} \right\}\neq \left\{0,0 \right\}, \left\{1,0 \right\}$.
\item For $|{\bf m }|\omega=1$, with
  the constraint ${\bf n}\neq 0$.
\item The sphere case ($\omega=1$), with the constraint $\left\{{\bf n},{\bf m}
  \right\}\neq \left\{0,0 \right\}, \left\{1,0 \right\}, \left\{0,\pm
    1 \right\}$.
This result
  is in agreement with that obtained by using the Wigner functions
  \cite{sphere}.  
\end{itemize}
When (\ref{muT/V}) is true a  derivative relation between
the 2D tensor and 2D vector wave functions holds: 
\be D_{\pm} (f^{\pm}_{{\bf n}{\bf m}}(\theta) e^{i {\bf
    m}\varphi})=c_{\T\,{\bf n}{\bf m}}f^{\pm \pm}_{{\bf n}{\bf
    m}}(\theta) e^{i {\bf m}\varphi}, \label{derivrel2}\ee 
where $c_{\T\,{\bf n}{\bf m}}$ are normalization constants which,
having chosen a convenient normalization for 
the wave functions,  can be fixed to be
$c_{\T\,{\bf n \, m}}= \sqrt{\mu^2_{\T\, {\bf n \, m}}}/\sqrt{2}$.   

It remains to expand the 6D fields in the action (\ref{spin-0-lc})
into their harmonics on the rugbyball and integrate over the extra
dimensions.  Thanks to the derivative relations
(\ref{derivrel},\ref{derivrel2}), and $F^2=const$, the 
mass-squared operator reduces to an algebraic matrix with constant
entries, which can easily be diagonalized.  We note that the
mass-matrix turns out to be
well-defined despite the singularities mentioned in
Subsection \ref{spin-0-action}. 

We end with the resulting spectrum for spin-0 fields (which can be
trusted when Eq. (\ref{muT/V}) holds). For definiteness we
focus here on the 6D supergravity setup, but there are no problems in deriving the squared
masses in the EYM$\Lambda$ case as well. We split the spectrum
according to the values of $l\equiv {\bf n}+|{\bf m }| \omega$:  
\begin{itemize}
 \item For $l=0$ $$ \frac{r_0^2}{4} M^2=0,\, 0,\, 2,\, [2]$$
\item For $l=1$ $$  \frac{r_0^2}{4} M^2= 2, \, 6, \, [2],\,[2],\,[2],\, [6]$$
\item For $l>1$ \bea M^2&=& \mu^2_{{\bf n \, m}} \quad \mbox{with
    multiplicity}\,\,\,  1[+3] \nonumber \\ M^2&=&\frac{4}{r_0^2}
  \left(1+\frac{r_0^2}{4}\mu^2_{{\bf n \, m}} - \sqrt{1+r_0^2
      \mu^2_{{\bf n \, m}}} \right) \,\, \mbox{with multiplicity}
  \,\,\,1[+1] \nonumber \\  M^2&=&
  \frac{4}{r_0^2}\left(1+\frac{r_0^2}{4}\mu^2_{{\bf n \, m}} +
    \sqrt{1+r_0^2 \mu^2_{{\bf n \, m}}} \right) \,\, \mbox{with
    multiplicity} \,\,\,1[+1] \nonumber\eea  
\end{itemize}
 where the square brackets denote helicity-0 components of higher spin
 fields and the remaining modes are physical spin-0 fields. We observe
 that there are neither ghosts nor tachyons and we recover the two
 massless fields discussed in Subsection \ref{spin06D}.   

As an effective check of the above spectrum we observe that
it correctly reduces, when $\omega=1$, to the sphere result
obtained by directly expanding the bulk fields over the Wigner functions
\cite{Salvio:2007mb}.

\end{document}